\documentclass[aps,reprint,showpacs,superscriptaddress]{revtex4-1}
\usepackage{graphicx} 
\usepackage{tabularx}
\usepackage{dcolumn} 
\usepackage{color}
\usepackage{bm}
\usepackage{amsmath}
\usepackage{amssymb}
\usepackage{multirow}
\newcommand{\bi}[1]{\ensuremath{\boldsymbol{#1}}} 
\newcommand{\nn}{\nonumber}
 
\begin{document} 
 
\title{ 
Symmetry protected vortex bound state in superfluid $^3$He B-phase
}

\author{Yasumasa Tsutsumi} 
\affiliation{Condensed Matter Theory Laboratory, RIKEN, 
Wako, Saitama 351-0198, Japan} 
\author{Takuto Kawakami} 
\affiliation{International Center for Materials Nanoarchitectonics (WPI-MANA) National Institute for Materials Science, 
Tsukuba 305-0044, Japan} 
\author{Ken Shiozaki}
\affiliation{Department of Physics, Kyoto University,
Kyoto 606-8502, Japan}
\author{Masatoshi Sato}
\affiliation{Department of Applied Physics, Nagoya University,
Nagoya 464-8603, Japan}
\author{Kazushige Machida} 
\affiliation{Department of Physics, Okayama University, 
Okayama 700-8530, Japan} 
\date{\today}

\begin{abstract} 
The superfluid $^3$He formed by spin-triplet $p$-wave Cooper pairs is a typical topological superfluid.
In the superfluid $^3$He B-phase, several kinds of vortices classified by spatial symmetries $P_1$, $P_2$, and $P_3$ are produced, where $P_1$ is inversion symmetry, $P_2$ is magnetic reflection symmetry, and $P_3$ is magnetic $\pi$-rotation symmetry.
We have calculated the vortex bound states by the Bogoliubov-de Gennes theory and the quasiclassical Eilenberger theory, and also clarified symmetry protection of the low energy excitations by the spatial symmetries.
On the symmetry protection, $P_3$ symmetry plays a key role which gives two-fold degenerate Majorana zero modes.
Then, the bound states in the most symmetric $o$ vortex with $P_1$, $P_2$, and $P_3$ symmetries and in $w$ vortex with $P_3$ symmetry have the symmetry protected degenerate Majorana zero modes.
On the other hand, zero energy modes in $v$ vortex, which is believed to be realized in the actual B-phase, are not protected, and in consequence become gapped by breaking axial symmetry.
The excitation gap may have been observed as the variation of critical velocity.
We have also suggested an experimental setup to create $o$ vortex with Majorana zero modes by a confinement and a magnetic field.
\end{abstract} 
 
 
 
\maketitle 

\section{Introduction}

Superfluid $^3$He is a condensate of spin-triplet $p$-wave Cooper pairs, which undergoes complex symmetry breaking, in addition to $U(1)_N$ gauge symmetry breaking~\cite{vollhardt:book}.
Without magnetic fields, the superfluid phases of $^3$He consist of the nodal gapped A-phase in a high-temperature and high-pressure region and the B-phase with an isotropic gap in another wide parameter region.
In the full gapped B-phase, the residual symmetry group is
\begin{align}
H_{\rm B}=SO(3)_{{\bm L}+{\bm S}}\times T\times C,
\end{align}
where $SO(3)_{{\bm L}+{\bm S}}$ is simultaneous rotation symmetry of orbital and spin spaces, $T$ is time-reversal symmetry, and $C$ is particle-hole symmetry.
Three-dimensional spin-triplet superfluids with $T$ and $C$, such as the B-phase, are topological phases belonging to class DIII in the Altland-Zrinbauer symmetry classes~\cite{schnyder:2008}.
In consequence of the topological order, the Majorana bound states and helical spin current exist on surfaces of the superfluid $^3$He B-phase~\cite{chung:2009,volovik:2009a,nagato:2009,tsutsumi:2011b,tsutsumi:2012c,wu:2013,okuda:2012}.

The topological phases are classified by whether there are discrete symmetries $T$ and $C$~\cite{schnyder:2008}.
It has been known that also spatial discrete symmetries in point group can give rise to a nontrivial topology of excitations in the topological phases~\cite{fu:2011}.
The topological excitations without a gap are robust if the symmetries are preserved.
Host matters of the symmetry protected excitations are called topological crystalline insulators~\cite{fu:2011} or topological crystalline superconductors~\cite{ueno:2013,tsutsumi:2013}.
The symmetry protected excitations are mainly discussed on gapless boundary states with reflection symmetries~\cite{chiu:2013,morimoto:2013}.
Recently, the topological classification is generalized for excitations in a defect with a two-fold symmetry including magnetic point group symmetry~\cite{shiozaki:2014}.

An example of the symmetry protected excitation in the superfluid $^3$He B-phase is the surface Majorana bound state under magnetic fields~\cite{mizushima:arXiv}.
The gapless Majorana bound state is preserved under weak magnetic fields parallel to the surface although time-reversal symmetry is broken~\cite{mizushima:2012b,mizushima:2012}.
This system has magnetic $\pi$-rotation symmetry instead of time-reversal symmetry because the flipped magnetic field by time-reversal operation is recovered by the $\pi$-rotation around the axis perpendicular to the surface.
The gapless Majorana bound state under magnetic fields is due to the protection by the magnetic $\pi$-rotation symmetry.

The symmetry protected Majorana excitations are also expected in the vortex bound state, where time-reversal symmetry is broken by the phase winding of the vortex.
For an axisymmetric vortex in the B-phase, the residual symmetry group is
\begin{align}
H_{\rm vortex}=U(1)_Q\times P\times C,
\label{eq:sym_vortex}
\end{align}
where the generator of the $U(1)_Q$ symmetry is $Q=L_z+S_z-N/2$ for a singly quantized vortex along the $z$-axis~\cite{salomaa:1987,volovik:book}.
The residual spatial symmetry $P$ is $P_2\times P_3$ for the most symmetric vortex called $o$ vortex~\cite{ohmi:1983}, where $P_2$ is magnetic reflection symmetry on a plane including the vortex line and $P_3$ is magnetic $\pi$-rotation symmetry around a axis perpendicular to the vortex line.
The combination of $P_2$ and $P_3$ gives inversion symmetry $P_1=P_2P_3$.
A vortex with a symmetry $P=P_1$, $P_2$, or $P_3$ is called $u$, $v$, or $w$ vortex, respectively, and the lowest symmetric vortex without the symmetry $P$ is called $uvw$ vortex.

In experiment, two types of vortices, V1 vortex in a high-temperature and high-pressure region and V2 vortex in a low-temperature and low-pressure region, were observed within the B-phase~\cite{volovik:book}.
The transition line between the vortex states was determined by the NMR measurement~\cite{hakonen:1983,hakonen:1983b,pekola:1984} and the measurement of critical velocity~\cite{pekola:1984b,pekola:1985}.
Since the transition pressure on the superfluid critical temperature is reproduced by the Ginzburg-Landau theory, V1 vortex and V2 vortex are believed to be axisymmetric $v$ vortex and non-axisymmetric $v$ vortex, respectively~\cite{thuneberg:1986b,thuneberg:1987}.
Breaking of the axial symmetry over the transition from V1 vortex to V2 vortex was confirmed by the observation of the Goldstone mode related to twisting of the anisotropic vortex core~\cite{kondo:1991}.
The existence of $P_2$ symmetry which is a direct evidence to form $v$ vortex, however, has not been verified.

For the identification of the kind of vortex in the superfluid $^3$He B-phase, the symmetry protected vortex bound state can be utilized.
Indeed, we will demonstrate that $P_3$ symmetry guarantees the existence of Majorana zero energy modes in the vortex bound state.
On the other hand, $P_2$ symmetry does not provide symmetry protected excitations in a vortex.
Therefore, many zero energy modes in axisymmetric $v$ vortex~\cite{silaev:2009} are gapped out by the axial symmetry breaking.
The formation of the gap in low energy excitations through the vortex transition implies the realization of $v$ vortex with only $P_2$ symmetry.

This paper is arranged as follows: In Sec.~II, we formulate the Bogoliubov-de Gennes (BdG) theory and the quasiclassical theory.
Vortex bound states are calculated by the BdG theory with order parameters (OPs) which are self-consistently obtained by the quasiclassical theory.
The possible spatial symmetries $P_1$, $P_2$, and $P_3$ for an axisymmetric vortex in the B-phase are summarized in Sec.~III.
The spatial symmetries play an important role in the topological classification of the vortex bound state.
In Sec.~IV, we discuss the bound state in the most symmetric $o$ vortex with $P_1$, $P_2$, and $P_3$ symmetries.
The vortex bound state has symmetry protected Majorana zero modes; however, $o$ vortex is not realized in the actual B-phase.
The $v$ vortex with $P_2$ symmetry is believed to be realized in the B-phase, whose bound state is discussed in Sec.~V.
There is a difference between low energy excitations in axisymmetric $v$ vortex and non-axisymmetric $v$ vortex in consequence of that $P_2$ symmetry does not protect zero energy modes.
We also discuss the bound state in $w$ vortex with $P_3$ symmetry in Sec.~VI.
Concerning the OP, the difference between $w$ vortex and $v$ vortex is only a phase of induced components of the OP around the vortex.
However, the bound state in $w$ vortex dramatically changes from that in $v$ vortex and has symmetry protected Majorana zero modes owing to $P_3$ symmetry.
We devote the final section to the summary in which we also mention topologically trivial bound states in $uvw$ vortex and $u$ vortex.

\section{Formulation}

\subsection{Bogoliubov-de Gennes theory}

We have numerically calculated the vortex bound states for some kinds of vortices in the B-phase by the BdG theory.
The BdG theory gives discretized modes in the vortex bound state at $\Delta^2/E_{\rm F}$ intervals, where $\Delta$ is a superfluid gap and $E_{\rm F}$ is the Fermi energy~\cite{caroli:1964}.
Note that the discreteness can be negligible when we consider physical quantities in the superfluid $^3$He owing to $\Delta/E_{\rm F}\sim 10^{-3}$~\cite{vollhardt:book};
however, we use the BdG theory in order to discuss the symmetry protection for the discretized modes in the vortex bound state.

The BdG equation in spin-triplet superfluid states is given as~\cite{kawakami:2011}
\begin{multline}
\int d{\bm r}_2
\begin{pmatrix}
\hat{h}({\bm r}_1,{\bm r}_2) & \hat{\Delta }({\bm r}_1,{\bm r}_2) \\
-\hat{\Delta }^{\dagger }({\bm r}_1,{\bm r}_2) & -\hat{h}^{\rm T}({\bm r}_1,{\bm r}_2)
\end{pmatrix}
\vec{u}_{\nu }({\bm r}_2)\\
=E_{\nu }\vec{u}_{\nu }({\bm r}_1),
\end{multline}
where the quasiparticle wave function with the $\nu$-th excited state with the eigenvalue $E_{\nu }$ is $\vec{u}_{\nu }({\bm r})=[u_{\nu }^{\uparrow }({\bm r}),u_{\nu }^{\downarrow }({\bm r}),v_{\nu }^{\uparrow }({\bm r}),v_{\nu }^{\downarrow }({\bm r})]^{\rm T}$.
The single particle Hamiltonian and the OP are described as
\begin{align}
\hat{h}({\bm r}_1,{\bm r}_2)=&\left[-\frac{\hbar^2\nabla_1^2}{2m}-E_{\rm F}\right]\delta({\bm r}_1-{\bm r}_2)\hat{1},\\
\hat{\Delta}({\bm r}_1,{\bm r}_2)
=&
\begin{pmatrix}
\Delta_{\uparrow\uparrow }({\bm r}_1,{\bm r}_2) & \Delta_{\uparrow\downarrow }({\bm r}_1,{\bm r}_2)\\
\Delta_{\downarrow\uparrow }({\bm r}_1,{\bm r}_2) & \Delta_{\downarrow\downarrow }({\bm r}_1,{\bm r}_2)
\end{pmatrix},
\end{align}
where $m$ is mass of the condensed particle and $\Delta_{\sigma\sigma'}({\bm r}_1,{\bm r}_2)=V(r')\sum_{\nu }u_{\nu }^{\sigma }({\bm r}_1)[v_{\nu }^{\sigma'}({\bm r}_2)]^*f(E_{\nu })$ with the interparticle interaction $V(r')$ at $r'=|{\bm r}_1-{\bm r}_2|$ and the Fermi distribution function $f(E_{\nu })$.

Here, we consider the bound state in a vortex along the $z$-axis; then the wave number $k_z$ becomes a well-defined quantum number.
The BdG equation for the $k_z$-resolved two-dimensional (2D) form is given as~\cite{kaneko:2012}
\begin{multline}
\int d{\bm \rho }_2
\begin{pmatrix}
\hat{h}_{k_z}(\bi{\rho }_1,\bi{\rho }_2) & \hat{\Delta }_{k_z}(\bi{\rho }_1,\bi{\rho }_2) \\
-\hat{\Delta }_{-k_z}^{\dagger }(\bi{\rho }_1,\bi{\rho }_2) & -\hat{h}_{-k_z}^{\rm T}(\bi{\rho }_1,\bi{\rho }_2)
\end{pmatrix}
\vec{u}_{\nu,k_z}(\bi{\rho }_2)\\
=E_{\nu,k_z}\vec{u}_{\nu,k_z}(\bi{\rho }_1),
\label{2DBdG}
\end{multline}
where
\begin{align}
\hat{h}_{k_z}(\bi{\rho }_1,\bi{\rho }_2)=\left[-\frac{\hbar^2\nabla_{\rm 2D}^2}{2m}-E_{\rm F}^{2{\rm D}}(k_z)\right]\delta(\bi{\rho }_1\!-\!\bi{\rho }_2)\hat{1},
\end{align}
with $\nabla_{\rm 2D}^2=\partial_{x_1}^2+\partial_{y_1}^2$.
The 2D form of the Fermi energy $E_{\rm F}^{2{\rm D}}(k_z)=(\hbar^2/2m)(k_{\rm F}^2-k_z^2)$ reflects the $k_z$-cross section of the Fermi surface, where $k_{\rm F}$ is the Fermi wave number.
Although the OP $\hat{\Delta }_{k_z}(\bi{\rho }_1,\bi{\rho }_2)$ should be calculated self-consistently with the quasiparticle wave function $\vec{u}_{\nu,k_z}({\bm \rho_i})$ and the eigenvalue $E_{\nu,k_z}$, an approximate solution can be derived from the quasiclassical theory for superfluid states with a small gap $\Delta/E_{\rm F}\ll 1$.
We expand the OP to the Fourier integral with the relative coordinate ${\bm \rho }'={\bm \rho }_1-{\bm \rho }_2$ as
\begin{align}\label{fourier}
\hat{\Delta }_{k_z}(\bi{\rho }_1,\bi{\rho }_2)=\int\frac{d\bi{k}^{\rm 2D}}{(2\pi)^2}e^{i\bi{k}^{\rm 2D}\cdot\bi{\rho }'}\hat{\Delta }(\bi{k},\bi{\rho })\Gamma(k),
\end{align}
where ${\bm \rho }=({\bm \rho }_1+{\bm \rho }_2)/2$ is the center-of-mass coordinate and ${\bm k}^{\rm 2D}=(k_x,k_y)$ is the 2D component of the relative momentum ${\bm k}$.
In this model, the Gaussian factor $\Gamma(k)\!=\!e^{-(k^2-k_{\rm F}^2)\xi_{\rm p}^2}$ indicates that the pairing interaction is non-zero near the Fermi surface $k=k_{\rm F}$ with the range of the interaction $\xi_{\rm p}$~\cite{mizushima:2010b}.
The OP $\hat{\Delta }({\bm k},{\bm \rho })$ is obtained by the quasiclassical theory in Sec.~II B.

If the $k_z$-resolved quasiparticle wave function $\vec{u}_{\nu,k_z}=\left( u_{\nu,k_z}^{\uparrow }, u_{\nu,k_z}^{\downarrow }, v_{\nu,-k_z}^{\uparrow }, v_{\nu,-k_z}^{\downarrow }\right)^{\rm T}$ satisfies the condition $u_{\nu,k_z}^{\uparrow }=\left(v_{\nu,-k_z}^{\uparrow }\right)^*$ and $u_{\nu,k_z}^{\downarrow }=\left(v_{\nu,-k_z}^{\downarrow }\right)^*$, the creation and annihilation operators are equivalent for the zero energy quasiparticle with $k_z=0$~\cite{kawakami:2011,sato:2014}.
Then, the quasiparticle is a Majorana zero mode.

\subsection{Quasiclassical theory}

The spatial structure of OP with a vortex has been calculated by the quasiclassical theory, which is valid for superfluids and superconductors with $\Delta\ll E_F$, such as the superfluid $^3$He.
We have found self-consistent solutions of the OP $\hat{\Delta }(\bar{\bm k},{\bm \rho })$ with the quasiclassical Green's function $\widehat{g}(\bar{\bi{k}},\bi{\rho },\omega_n)$ by the Eilenberger equation~\cite{eilenberger:1968,serene:1983,fogelstrom:1995}
\begin{multline}
-i\hbar\bi{v}(\bar{\bi{k}})\cdot\bi{\nabla }\widehat{g}(\bar{\bi{k}},\bi{\rho },\omega_n)\\
= \left[
\begin{pmatrix}
i\omega_n\hat{1} & -\hat{\Delta }(\bar{\bi{k}},\bi{\rho }) \\
\hat{\Delta }^{\dagger }(\bar{\bi{k}},\bi{\rho }) & -i\omega_n\hat{1}
\end{pmatrix}
,\widehat{g}(\bar{\bi{k}},\bi{\rho },\omega_n) \right],
\label{Eilenberger eq}
\end{multline}
where the wide hat indicates the 4 $\times$ 4 matrix in particle-hole and spin spaces.
The quasiclassical Green's function is described in particle-hole space by
\begin{align}
\widehat{g}(\bar{\bi{k}},\bi{\rho },\omega_n) = -i\pi
\begin{pmatrix}
\hat{g}(\bar{\bi{k}},\bi{\rho },\omega_n) & i\hat{f}(\bar{\bi{k}},\bi{\rho },\omega_n) \\
-i\underline{\hat{f}}(\bar{\bi{k}},\bi{\rho },\omega_n) & -\underline{\hat{g}}(\bar{\bi{k}},\bi{\rho },\omega_n)
\end{pmatrix},
\end{align}
where $\omega_n=(2n+1)\pi k_B T$ is the Matsubara frequency and $\bar{\bm k}$ is the normalized relative momentum on the Fermi surface.
The quasiclassical Green's function satisfies the normalization condition $\widehat{g}^2=-\pi^2\widehat{1}$.
The Fermi velocity is given as $\bi{v}(\bar{\bi{k}})=v_F\bar{\bi{k}}$ on a three-dimensional Fermi sphere.

The spin-triplet OP is defined by
\begin{align}
\hat{\Delta }(\bar{\bi{k}},\bi{\rho })=i\bi{d}(\bar{\bi{k}},\bi{\rho })\cdot\hat{\bi{\sigma }}\hat{\sigma_y},
\label{OP}
\end{align}
where $\hat{\bi{\sigma }}$ is the Pauli matrix.
The $d$-vector is perpendicular to the spin $\bi{S}$ of a Cooper pair, namely, $\bi{d}\cdot\bi{S}=0$.
The description using projections of spin angular momentum is more convenient than the description by the $d$-vector for the OP with an axisymmetric vortex, namely,
\begin{align}
\hat{\Delta }(\bar{\bi{k}},\bi{\rho })=
\begin{pmatrix}
-\sqrt{2}C_+(\bar{\bi{k}},\bi{\rho }) & C_0(\bar{\bi{k}},\bi{\rho }) \\
C_0(\bar{\bi{k}},\bi{\rho }) & \sqrt{2}C_-(\bar{\bi{k}},\bi{\rho })
\end{pmatrix},
\end{align}
where $C_{\pm }=(d_x\mp id_y)/\sqrt{2}$ and $C_0=d_z$.
Each coefficient can be expanded in projections of orbital angular momentum,
\begin{multline}
C_a(\bar{\bi{k}},\bi{\rho })=C_{a+}(\bi{\rho })\bar{k}_+ + C_{a0}(\bi{\rho })\bar{k}_0 + C_{a-}(\bi{\rho })\bar{k}_-,
\label{eq:C}
\end{multline}
with $a=0$ or $\pm$, where $\bar{k}_{\pm }=(\bar{k}_x\pm i\bar{k}_y)/\sqrt{2}$ and $\bar{k}_0=\bar{k}_z$~\cite{salomaa:1987}.

The self-consistent condition for $\hat{\Delta }$ is given as
\begin{align}
\hat{\Delta }(\bar{\bi{k}},\bi{\rho }) = N_0\pi k_BT\sum_{|\omega_n| \le \omega_{\rm c}}\left\langle V(\bar{\bi{k}}, \bar{\bi{k}}') \hat{f}(\bar{\bi{k}}',\bi{\rho },\omega_n)\right\rangle_{\bar{\bi{k}}'},
\label{order parameter}
\end{align}
where $N_0$ is the density of states in the normal state,
$\omega_{\rm c}$ is a cutoff energy setting $\omega_{\rm c}=20k_B T_{\rm c}$ with the critical temperature $T_{\rm c}$,
and $\langle\cdots\rangle_{\bar{\bi{k}}}$ indicates the Fermi surface average.
The pairing interaction $V(\bar{\bi{k}}, \bar{\bi{k}}')=3g\bar{\bi{k}}\cdot\bar{\bi{k}}'$, where $g$ is a coupling constant with the relation $(gN_0)^{-1}=\ln(T/T_c)+\pi k_BT\sum_{|\omega_n| \le \omega_c}|\omega_n|^{-1}$.
We solve Eq.~\eqref{Eilenberger eq} and Eq.~\eqref{order parameter} alternately at $T=0.2T_{\rm c}$, and obtain a self-consistent solution.
Then, we use the self-consistent OP $\hat{\Delta }(\bar{\bm k},{\bm \rho })$ after the replacement of $\bar{\bm k}$ by ${\bm k}/k_{\rm F}$ as $\hat{\Delta }({\bm k},{\bm \rho })$ in Eq.~\eqref{fourier}.

\section{Spatial symmetries for an axisymmetric vortex}

For the OP with an axisymmetric vortex, the coefficients in Eq.~\eqref{eq:C} are described by $C_{ab}(\bm{\rho })=C_{ab}(\rho)e^{in_{ab}\phi}$, where $\rho$ is a radial distance from a vortex core and $\phi$ is an azimuthal angle in the $xy$-plane.
Since the OP has $U(1)_Q$ symmetry with $Q=L_z+S_z-N/2$ as Eq.~\eqref{eq:sym_vortex}, the phase winding number $n_{ab}$ in each coefficient satisfies
\begin{equation}
\begin{split}
n_{a+1,b}=&n_{a,b}-1,\\
n_{a,b+1}=&n_{a,b}-1.
\end{split}
\end{equation}
The following coefficients are finite in the bulk B-phase:
\begin{align}
C_{+-}=C_{00}=C_{-+}=\Delta_{\rm B},
\end{align}
where $\Delta_{\rm B}$ is the amplitude of a gap in the bulk, because the total angular momentum ${\bm J}={\bm L}+{\bm S}={\bm 0}$ in the B-phase~\cite{vollhardt:book}.
With a singly quantized vortex, the OP has $2\pi$-phase winding around the vortex, namely, phase winding numbers
\begin{align}
n_{+-}=n_{00}=n_{-+}=1.
\end{align}
Then, the general description of the coefficients with a singly quantized axisymmetric vortex in the superfluid $^3$He B-phase is
\begin{align}
C(\bi{\rho })=
\begin{pmatrix}
C_{++}(\rho)e^{-i\phi } & C_{+0}(\rho) & C_{+-}(\rho)e^{i\phi } \\
C_{0+}(\rho) & C_{00}(\rho)e^{i\phi } & C_{0-}(\rho)e^{2i\phi } \\
C_{-+}(\rho)e^{i\phi } & C_{-0}(\rho)e^{2i\phi } & C_{--}(\rho)e^{3i\phi }
\end{pmatrix},
\label{eq:axivortex}
\end{align}
where $C_{ab}$ approaches $\Delta_{\rm B}$ for $a+b=0$ and vanishes for $a+b\neq 0$ when $\rho\to\infty$.

The B-phase with a vortex can have the additional symmetry $P$ as Eq.~\eqref{eq:sym_vortex}.
Under the symmetry operation, the semi-classical BdG Hamiltonian
\begin{align}
\widehat{H}_{\rm BdG}({\bm k},{\bm \rho})=
\begin{pmatrix}
\hat{h}(\bm k) & \hat{\Delta }({\bm k},{\bm \rho}) \\
\hat{\Delta }^{\dagger }({\bm k},{\bm \rho}) & -\hat{h}^{\rm T}(-{\bm k})
\end{pmatrix},
\end{align}
satisfies
\begin{align}
\widehat{\mathcal{P}}\widehat{H}_{\rm BdG}({\bm k},{\bm \rho})\widehat{\mathcal{P}}^{-1}=\widehat{H}_{\rm BdG}({\bm k}',{\bm \rho}'),
\end{align}
where the symmetry operator is $\widehat{\mathcal{P}}={\rm diag}(\hat{\mathcal{P}},\hat{\mathcal{P}}^*)$ and ${\bm k}$ and ${\bm \rho}$ are transformed into ${\bm k}'$ and ${\bm \rho}'$, respectively, by the operation.
The normal state Hamiltonian $\hat{h}({\bm k})=(\hbar^2/2m)({\bm k}^2-k_{\rm F}^2)\hat{1}$ implies $\hat{\mathcal{P}}\hat{h}({\bm k})\hat{\mathcal{P}}^{-1}=\hat{h}({\bm k})$.
Since the operation $\hat{\mathcal{P}}\hat{\Delta }({\bm k},{\bm \rho})\left(\hat{\mathcal{P}}^*\right)^{-1}$ can be regarded as $\mathcal{P}C({\bm k},{\bm \rho})\left(\mathcal{P}^*\right)^{-1}$, the inversion symmetry operator $\mathcal{P}_1$ acts on $C_a({\bm k},\rho,\phi)$ as
\begin{multline}
\mathcal{P}_1C({\bm k},\rho,\phi)\left(\mathcal{P}_1^*\right)^{-1}=C(-{\bm k},\rho,\phi+\pi)\\
=\begin{pmatrix}
C_{++}(\rho)e^{-i\phi } & -C_{+0}(\rho) & C_{+-}(\rho)e^{i\phi } \\
-C_{0+}(\rho) & C_{00}(\rho)e^{i\phi } & -C_{0-}(\rho)e^{2i\phi } \\
C_{-+}(\rho)e^{i\phi } & -C_{-0}(\rho)e^{2i\phi } & C_{--}(\rho)e^{3i\phi }
\end{pmatrix}
\begin{pmatrix}
k_+ \\ k_0 \\ k_-
\end{pmatrix}.
\label{eq:p1sym}
\end{multline}
The symmetry $P_2$ is the combined symmetry with the time-reversal and mirror reflection on a plane including the vortex line, namely the magnetic reflection symmetry.
The time-reversal operator $\mathcal{T}$ acts as $\mathcal{T}C_{a,b}(\rho,\phi)\left(\mathcal{T}^*\right)^{-1}=-C_{-a,-b}^*(\rho,\phi)$ and the mirror reflection operator $\mathcal{M}$ acts as $\mathcal{M}C_{a,b}(\rho,\phi)\left(\mathcal{M}^*\right)^{-1}=-C_{-a,-b}(\rho,-\phi)$; therefore,
\begin{multline}
\mathcal{P}_2C({\bm k},\rho,\phi)\left(\mathcal{P}_2^*\right)^{-1}\\
=\begin{pmatrix}
C_{++}^*(\rho)e^{-i\phi } & C_{+0}^*(\rho) & C_{+-}^*(\rho)e^{i\phi } \\
C_{0+}^*(\rho) & C_{00}^*(\rho)e^{i\phi } & C_{0-}^*(\rho)e^{2i\phi } \\
C_{-+}^*(\rho)e^{i\phi } & C_{-0}^*(\rho)e^{2i\phi } & C_{--}^*(\rho)e^{3i\phi }
\end{pmatrix}
\begin{pmatrix}
k_+ \\ k_0 \\ k_-
\end{pmatrix}.
\label{eq:p2sym}
\end{multline}
The symmetry $P_3$ is the magnetic $\pi$-rotation symmetry around an axis perpendicular to the vortex line.
The $\pi$-rotation operator $\mathcal{R}$ acts as $\mathcal{R}C_{a,b}(\rho,\phi)\left(\mathcal{R}^*\right)^{-1}=C_{-a,-b}(\rho,\pi -\phi)$.
Then, the combination of $\mathcal{T}$ and $\mathcal{R}$ gives
\begin{multline}
\mathcal{P}_3C({\bm k},\rho,\phi)\left(\mathcal{P}_3^*\right)^{-1}\\
=\begin{pmatrix}
C_{++}^*(\rho)e^{-i\phi } & -C_{+0}^*(\rho) & C_{+-}^*(\rho)e^{i\phi } \\
-C_{0+}^*(\rho) & C_{00}^*(\rho)e^{i\phi } & -C_{0-}^*(\rho)e^{2i\phi } \\
C_{-+}^*(\rho)e^{i\phi } & -C_{-0}^*(\rho)e^{2i\phi } & C_{--}^*(\rho)e^{3i\phi }
\end{pmatrix}
\begin{pmatrix}
k_+ \\ k_0 \\ k_-
\end{pmatrix}.
\label{eq:p3sym}
\end{multline}

\section{$o$ vortex}
\label{sec:o_vortex}

The $o$ vortex is the most symmetric vortex in the superfluid $^3$He B-phase.
Since it has the all possible discrete symmetries $P_1$, $P_2$, and $P_3$, several coefficients should vanish as
\begin{align}
C(\bi{\rho })=
\begin{pmatrix}
C_{++}(\rho)e^{-i\phi } & 0 & C_{+-}(\rho)e^{i\phi } \\
0 & C_{00}(\rho)e^{i\phi } & 0 \\
C_{-+}(\rho)e^{i\phi } & 0 & C_{--}(\rho)e^{3i\phi }
\end{pmatrix},
\end{align}
where all remaining coefficients $C_{ab}(\rho)$ are real and they approach $C_{+-}=C_{00}=C_{-+}=\Delta_{\rm B}$ and $C_{++}=C_{--}=0$ when $\rho\rightarrow\infty$.
The component $C_{++}$ ($C_{--}$) is induced around the vortex by the spatial variation of the bulk component $C_{+-}$ ($C_{-+}$)~\cite{ohmi:1983}.

\begin{figure}
\begin{center}
\includegraphics[width=8.5cm]{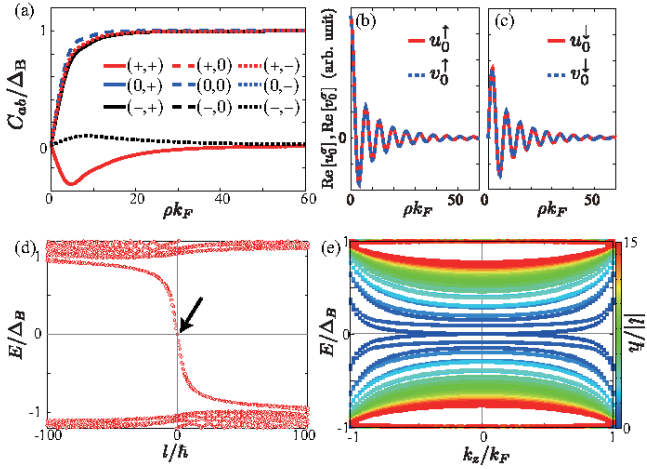}
\end{center}
\caption{\label{ovor}(Color online) Numerical results for $o$ vortex.
(a) OP configuration $C_{ab}(\rho)$ for finite components $C_{+-}$, $C_{00}$, $C_{-+}$, $C_{++}$, and $C_{--}$.
Real part of wave functions $u^{\sigma}_0(\rho)$ and $v^{\sigma}_0(\rho)$ with $E_{\nu,k_z}=0$ for the $\uparrow$-spin state (b) and $\downarrow$-spin state (c).
(d) Eigenvalues for quasiparticles with $k_z=0$. The degenerate eigenstates indicated by an arrow correspond to the two wave functions in (b) and (c).
(e) $k_z$-dispersion of the vortex bound state whose color denotes angular momentum $l$ of quasiparticles.
}
\end{figure}

The self-consistently obtained OP by the quasiclassical theory is shown in Fig.~\ref{ovor}(a).
The bulk components $C_{+-}$, $C_{00}$, and $C_{-+}$ rise up with $\rho$-linear from a vortex core, where $C_{00}$ recovers the bulk gap $\Delta_{\rm B}$ with slightly shorter length than $C_{+-}$ and $C_{-+}$.
Rises of $C_{++}$ and $C_{--}$ are $\rho$-linear and $\rho$-cubic, respectively, which reflect phase winding numbers $n_{++}=-1$ and $n_{--}=3$.
This difference makes a little variance between $C_{+-}$ and $C_{-+}$ via a coupling between the same orbital state.

The discretized eigenvalues of the vortex bound state, which is derived from the BdG theory with the self-consistent OP, are shown in Figs.~\ref{ovor}(d) and \ref{ovor}(e).
In Fig.~\ref{ovor}(d), the eigenvalues for $k_z=0$ are classified into quantized orbital angular momentum $l$.
The low energy eigenvalues are discretized at the order of $\Delta_{\rm B}^2/E_{\rm F}$ intervals, where we take $\Delta_{\rm B}/E_{\rm F}=0.1$.
The vortex bound state has spin degenerate exact two zero energy modes at $l=0$ indicated by an arrow.
The wave functions of the degenerate zero energy modes for $\uparrow$-spin and $\downarrow$-spin states are shown in Figs.~\ref{ovor}(b) and \ref{ovor}(c), respectively.
Since the orbital chirality of the $\uparrow$-spin state is antiparallel to the vorticity but the orbital chirality of the $\downarrow$-spin state is parallel to the vorticity, $u_0^{\uparrow }({\bm\rho})=\left[v_0^{\uparrow }({\bm\rho})\right]^*\propto J_0(k_{\rm F}\rho)\exp(-\rho/\xi)$ and $u_0^{\downarrow }({\bm\rho})=\left[v_0^{\downarrow }({\bm\rho})\right]^*\propto J_1(k_{\rm F}\rho)\exp(-\rho/\xi)e^{i\phi }$~\cite{gurarie:2007,mizushima:2010}, where $J_n$ is the $n$-th order Bessel function and $(k_{\rm F}\xi)^{-1}=\Delta_{\rm B}/E_{\rm F}=0.1$.
These zero energy modes are Majorana zero modes because the quasiparticle wave functions for both spin states satisfy the condition $u_0^{\uparrow }=\left(v_0^{\uparrow }\right)^*$ and $u_0^{\downarrow }=\left(v_0^{\downarrow }\right)^*$, as demonstrated in Fig.~\ref{ovor}(b) and \ref{ovor}(c).

The $k_z$-dispersion of the vortex bound state for various $l$'s is shown in Fig.~\ref{ovor}(e), where two different spin states exist for each $l$.
As demonstrated in Appendix~\ref{app:ovortex}, the $l=0$ branches linearly cross the zero energy at $k_z=0$ as $E_{\pm}\propto \pm k_z$.
Their wave functions are distinguished into $\vec{u}_{\pm}=\left(\vec{u}_0^{\uparrow}\pm\vec{u}_0^{\downarrow}\right)/\sqrt{2} + \vec{\mathcal{O}}(k_z/k_{\rm F})$, where $\vec{u}_0^{\uparrow}=\left(u_0^{\uparrow},0,\left[u_0^{\uparrow}\right]^*,0\right)^{\rm T}$ and $\vec{u}_0^{\downarrow}=\left(0,u_0^{\downarrow},0,\left[u_0^{\downarrow}\right]^*\right)^{\rm T}$ are the wave functions of the Majorana zero modes in the $\uparrow$-spin and $\downarrow$-spin states, respectively.
Note that the inversion symmetry $P_1$ maps each eigenstate to another one, namely, $\vec{u}_{\pm}=\widehat{\mathcal{P}}_1\vec{u}_{\mp}$.

The spin degenerate Majorana zero modes are protected by spatial symmetry.
The vortex bound state with $P_2$ and $P_3$ symmetries is classified by a topological invariant $\mathbb{Z}$ which is demonstrated in Appendix~\ref{app:clifford} by using Clifford algebras.
Note that $P_1$ symmetry does not provide symmetry protected defect zero modes in general.

The appropriate topological invariant for $o$ vortex is calculated by
the one-dimensional winding number, which is evaluated from the BdG
Hamiltonian at an infinite point from a vortex 
\begin{align}
\widehat{H}_{\rm BdG}({\bm k},\phi)=
\begin{pmatrix}
\hat{h}({\bm k}) & \hat{\Delta }({\bm k},\phi) \\
\hat{\Delta }^{\dagger }({\bm k},\phi) & -\hat{h}^{\rm T}(-{\bm k})
\end{pmatrix},
\end{align}
where $\hat{h}({\bm k})=(\hbar^2/2m)({\bm k}^2-k_{\rm F}^2)\hat{1}$ and
\begin{align}
\hat{\Delta }({\bm k},\phi)=\frac{\Delta_{\rm B}}{k_{\rm F}}
\begin{pmatrix}
-k_x+ik_y & k_z \\
k_z & k_x+ik_y
\end{pmatrix}e^{i\phi }.\nn
\end{align}
From combination of $P_3$ and particle-hole symmetry, 
the BdG Hamiltonian for $o$ vortex obeys
$(\widehat{\mathcal{C}}\widehat{\mathcal{P}}_3)\widehat{H}_{\rm
BdG}(k_x,-k_y,-k_z,-\phi)(\widehat{\mathcal{C}}\widehat{\mathcal{P}}_3)^{-1}=-\widehat{H}_{\rm
BdG}({\bm k},\phi)$ where $\widehat{\mathcal{P}}_3=i\hat{\sigma}_z\widehat{\tau}_z\mathcal{K}$ and $\widehat{\mathcal{C}}=\widehat{\tau
}_x\mathcal{K}$ are operators for the magnetic $\pi$-rotation and
particle-hole symmetry with the complex conjugation operator
$\mathcal{K}$.  
The combined symmetry defines the chiral symmetry, 
$\widehat{\Gamma }\widehat{H}_{\rm
BdG}(k_x,k_y=0,k_z=0,\phi=0,\pi)\widehat{\Gamma }^{-1}=-\widehat{H}_{\rm
BdG}(k_x,k_y=0,k_z=0,\phi=0,\pi)$ with $\widehat{\Gamma }=\widehat{\tau
}_y\hat{\sigma }_z$, in the symmetric space $k_y=k_z=0$
and $\phi=0$ or $\pi$.
The chiral symmetry enables us to introduce the one-dimensional winding number as~\cite{sato:2009c,sato:2011}
\begin{widetext}
\begin{align}
w^{\phi=0,\pi }=-\frac{1}{4\pi i}\int dk_x{\rm tr}\left[\widehat{\Gamma }\widehat{H}^{-1}_{\rm BdG}(k_x,k_y=0,k_z=0,\phi=0,\pi)\partial_{k_x}\widehat{H}_{\rm BdG}(k_x,k_y=0,k_z=0,\phi=0,\pi)\right].
\label{eq:1dwinding}
\end{align}
\end{widetext}
For $o$ vortex, the one-dimensional winding number is evaluated as
$w^0=2$ and $w^{\pi }=-2$. 
The difference of the winding $(w^0-w^{\pi })/2=2$ provides
the $\mathbb{Z}$ topological invariant for $o$ vortex.
This topological number corresponds to the number of the zero energy states at
$k_z=0$. (More precisely, the $\mathbb{Z}$ topological invariant is
equal to the index ${\rm tr}\widehat{\Gamma}$ of the quasiparticle states
at $k_z=0$. Since the chiral symmetry requires that the zero energy states at
$k_z=0$ are eigenstates of $\widehat{\Gamma}$,  the index ${\rm
tr}\widehat{\Gamma}$ reduces to the difference between the number of
the zero energy states with eigenvalue $\widehat{\Gamma}=+1$ and that
with $\widehat{\Gamma}=-1$~\cite{sato:2011}. Hence, if
$(w^0-w^{\pi})/2=N$, there exist at least $|N|$ zero energy states at $k_z=0$.) 
The obtained topological number, i.e. $(w^0-w^\pi)/2=2$, guarantees the
existence of two zero energy states at $k_z=0$, which are indeed realized
as the two $l=0$ zero modes in Fig.~\ref{ovor}.

Note that the second Chern number, which characterizes bound states in a line defect for the symmetry class D, vanishes in the presence of $P_1$ symmetry.
The second Chern number ${\rm Ch}_2$ is obtained by~\cite{qi:2008, teo:2010}
\begin{align}
{\rm Ch}_2=\frac{1}{8\pi^2}\int d{\bm k}d\phi\epsilon^{ijk}{\rm tr}[f_{\phi i}f_{jk}],
\label{eq:secondCh}
\end{align}
where $f_{\alpha\beta}^{mn}=\partial_{\alpha}a_{\beta}^{mn}
-\partial_{\beta}a_{\alpha}^{mn}+i[a_{\alpha},a_{\beta}]^{mn}$ is the
curvature of the non-Abelian Berry connection $a_\alpha$ with
$\alpha=(\phi, i)=(\phi,k_x,k_y,k_z)$, and 
the non-Abelian Berry connection is given by
\begin{align}
a_{\alpha}^{mn}=-i\langle m,\bi{k},\phi|\partial_{\alpha}|n,\bi{k},\phi\rangle,
\end{align}
with eigenstates $|m, \bi{k}, \phi\rangle$ and $|n, \bi{k}, \phi\rangle$ 
of the BdG Hamiltonian $\widehat{H}_{\rm BdG}({\bm k},\phi)$.
The indices $m$ and $n$ label quasiparticle states with negative energies.
For $o$ vortex, the eigenstates have $P_1$ symmetry which gives the periodicity of Berry curvature, $f_{\phi i}({\bm k},\phi+\pi)=-f_{\phi i}(-{\bm k},\phi)$ and $f_{jk}({\bm k},\phi+\pi)=f_{jk}(-{\bm k},\phi)$.
Therefore, the integral in Eq.~\eqref{eq:secondCh} yields ${\rm Ch}_2=0$.
Since the second Chern number generally vanishes when there is $P_1$ symmetry, it is not appropriate for the topological number of the bound states in $o$ vortex.


\section{$v$ vortex}

In the actual B-phase, it is believed that two kinds of $v$ vortices are realized.
One $v$ vortex has axial symmetry and the other $v$ vortex breaks the axial symmetry.
In this section, we show the difference of the vortex bound states between the two kinds of $v$ vortices.

\subsection{Axisymmetric $v$ vortex}

The axisymmetric $v$ vortex has the magnetic reflection symmetry $P_2$.
From Eq.~\eqref{eq:p2sym}, all coefficients $C_{ab}(\rho)$ in Eq.~\eqref{eq:axivortex} are real, where $C_{+-}=C_{00}=C_{-+}=\Delta_{\rm B}$ and the other coefficients vanish when $\rho\rightarrow\infty$.
Since $C_{0+}$ and $C_{+0}$ corresponding to the A-phase and $\beta$-phase components, respectively, do not have any phase winding, they can compensate a vortex core.

\begin{figure}
\begin{center}
\includegraphics[width=8.5cm]{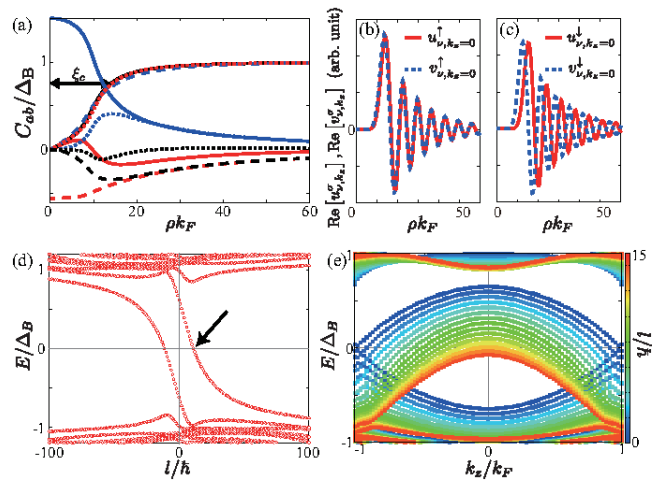}
\end{center}
\caption{\label{vvor}(Color online) Numerical results for axisymmetric $v$ vortex.
(a) OP configuration $C_{ab}(\rho)$ whose legends are identical to that in Fig.~\ref{ovor}(a).
(b) and (c) Real part of wave function $u^{\sigma}_{\nu,k_z=0}(\rho)$ and $v^{\sigma}_{\nu,k_z=0}(\rho)$ for a state with $E_{\nu,k_z=0}\approx0$.
(d) Eigenvalues for quasiparticles with $k_z=0$. The eigenstate indicated by an arrow corresponds to the wave function in (b) and (c).
(e) $k_z$-dispersion of quasiparticles with $l\ge0$ whose color denotes angular momentum $l$ of quasiparticles.
}
\end{figure}

The self-consistently obtained OP by the quasiclassical theory is shown in Fig.~\ref{vvor}(a).
The A-phase component $C_{0+}$ compensates the vortex core with larger amplitude than that of the bulk components in the B-phase.
The $\beta$-phase component $C_{+0}$ also compensates the vortex core; however, the amplitude is smaller than that of $C_{0+}$ and the sign is opposite.
Away from the vortex core, $C_{0+}$ and $C_{+0}$ are identical to $C_{0-}$ and $C_{-0}$, respectively, in order that total angular momentum becomes zero.
A core radius $\xi_{\rm c}$ of the $v$ vortex, which is characterized by the healing length of the bulk components, is longer than that of $o$ vortex because a loss of the condensation energy at the $v$ vortex core is small.
Other $C_{++}$ and $C_{--}$ components are also induced slightly, which change the sign away from the vortex core.

The discretized eigenvalues derived from the BdG equation by using the self-consistent OP are shown in Figs.~\ref{vvor}(d) and \ref{vvor}(e).
In Fig.~\ref{vvor}(d), the eigenvalues for $k_z=0$ are classified into quantized orbital angular momentum $l$ along the vortex line.
The quasiparticles with $l=0$ and $k_z=0$ have the finite energy by the $\pi/2$-phase shift owing to the A-phase component compensating the vortex core [see Appendix~\ref{app:shift}].
Discretized eigenvalues for $k_z=0$ may be situated on the zero energy with finite $\pm l$, accidentally, according to a core radius.
The wave function of an accidental zero energy mode, indicated by an arrow in Fig.~\ref{vvor}(d), is shown in Figs.~\ref{vvor}(b) and \ref{vvor}(c) for $\uparrow$-spin and $\downarrow$-spin parts, respectively.
The quasiparticle wave function satisfies $u_{\nu,k_z=0}^{\uparrow }=\left(v_{\nu,k_z=0}^{\uparrow }\right)^*$ while $u_{\nu,k_z=0}^{\downarrow }\ne\left(v_{\nu,k_z=0}^{\downarrow }\right)^*$;
therefore, the quasiparticle is not the Majorana zero mode~\cite{kawakami:2011,sato:2014}.
It is a natural consequence because the particle-hole operation changes the sign of orbital angular momentum.

The $k_z$-dispersion of the vortex bound state for $l\ge 0$ is shown in Fig.~\ref{vvor}(e).
Eigenvalues for $l'=-l\le 0$ are given by $E'=-E$ with $k_z'=-k_z$ owing to the particle-hole symmetry.
The eigenvalues with $l=0$ approach the zero energy at $k_z=\pm k_F$ owing to an induced $C_{-0}$ component around the vortex, which is different from the result by Silaev~\cite{silaev:2009} [see Appendix~\ref{app:shift}].
Since momentum $k_z$ is a continuous quantity, the eigenvalues with $|l|\lesssim\xi_{\rm c}k_{\rm F}$ cross the zero energy at finite $k_z$.
The zero energy modes with finite $k_z$ are, however, also not Majorana zero modes because signs of $l$ and $k_z$ for the zero energy modes are changed by the particle-hole operation, that is, the zero energy modes are mapped to different zero energy modes by the particle-hole operation.

The trivial vortex bound state without Majorana zero modes can be understood by the topological arguments in Appendix~\ref{app:clifford}.
The vortex bound states without any symmetry protections are clearly shown as gaps of low energy excitations in non-axisymmetric $v$ vortex which breaks axial symmetry but keeps $P_2$ symmetry.

\subsection{Non-axisymmetric $v$ vortex}

\begin{figure}
\begin{center}
\includegraphics[width=8.5cm]{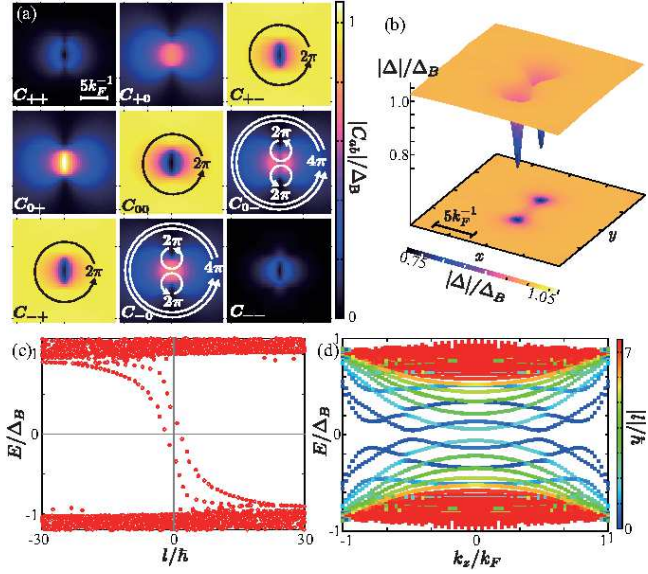}
\end{center}
\caption{\label{dvor}(Color online) Numerical results for non-axisymmetric $v$ vortex.
2D profile for each OP component $C_{ab}(\bm{\rho})$ (a) and the root-mean-square value of the gap, $|\Delta|=\sqrt{\langle{\rm Tr}[\hat{\Delta }\hat{\Delta }^{\dagger }]\rangle_{\bar{\bi{k}}}/2}$, (b).
(c) Eigenvalues for quasiparticles with $k_z=0$.
(d) $k_z$-dispersion of the vortex bound state whose color denotes angular momentum $l$ of quasiparticles.
}
\end{figure}

The self-consistently obtained OP for non-axisymmetric $v$ vortex is shown in Figs.~\ref{dvor}(a) and \ref{dvor}(b).
Far away from the vortex core, non-axisymmetric $v$ vortex is also described by the coefficients in Eq.~\eqref{eq:axivortex}; however, the original $4\pi$-phase singularity of $C_{-0}$ and $C_{0-}$ is split into two singularities with the $2\pi$-phase winding as shown in Fig.~\ref{dvor}(a).
This deformation breaks $U(1)_Q$ symmetry but keeps $P_2$ symmetry.
In Fig.~\ref{dvor}(b), we show the root-mean-square value of the gap, $|\Delta|=\sqrt{\langle{\rm Tr}[\hat{\Delta }\hat{\Delta }^{\dagger }]\rangle_{\bar{\bi{k}}}/2}$.
Since four components $C_{0+}$, $C_{+0}$, $C_{-0}$, and $C_{0-}$ are finite at the vortex center, substantial gap opens on the vortex center.
In return for the gap on the vortex center, finite minima of $|\Delta|$ appear on the phase singularity of $C_{-0}$ and $C_{0-}$.

The discretized eigenvalues derived from the BdG equation by using the self-consistent OP are shown in Figs.~\ref{dvor}(c) and \ref{dvor}(d).
In Fig.~\ref{dvor}(c), the eigenvalues for $k_z=0$ are classified into orbital angular momentum $l$ along the vortex line.
The $k_z$-dispersion of the vortex bound state with values of $l$ is shown in Fig.~\ref{dvor}(d).
Note that $l$ is not a quantum number but it is calculated by
\begin{align}
l=-i\hbar\int d\bi{\rho }\vec{u}_{\nu,k_z}^{\dagger }({\bm \rho})\partial_{\phi }\vec{u}_{\nu,k_z}({\bm \rho}),
\label{eq:ang_moment}
\end{align}
for each eigenstate labeled $\nu$ and $k_z$.
The quasiparticle excitations in the vortex bound state have a gap by the hybridization of different $l$ eigenstates as discussed below.
This is a consequence of that $P_2$ symmetry does not protect zero energy modes in a line defect.
The amplitude of the excitation gap is of the order of $\Delta/E_{\rm F}$.
Therefore, the gap becomes larger when $k_z$ approach $k_{\rm F}$ because the effective Fermi energy $E_{\rm F}^{\rm 2D}=(\hbar^2/2m)(k_{\rm F}^2-k_z^2)$ decreases.

Here, let us consider the hybridization of different $l$ eigenstates for non-axisymmetric $v$ vortex.
For axisymmetric $v$ vortex, since angular momentum $l$ is a well-defined quantum number, different $l$ eigenstates do not hybridize each other.
The reduction of axial symmetry to $n$-fold rotational symmetry, however, gives hybridization between the states with angular momentum $l$ and $l+nm$, where $m\in\mathbb{Z}$.
Then, $l$ and $l+2m$ eigenstates for non-axisymmetric $v$ vortex, which has two-fold rotational symmetry, are hybridized.
For the $k_z$-dispersion of the original $v$ vortex bound state, branches of $l$ and $-l$ eigenstates are crossed on the zero energy, where $l\ge 0$ eigenstates are shown in Fig.~\ref{vvor}(e) and $l'=-l$ eigenstates have eigenvalues $E'=-E$.
Since the $l$ and $-l$ eigenstates are hybridized for non-axisymmetric $v$ vortex, the quasiparticle excitations have a gap.

\section{$w$ vortex}
\label{sec:w_vortex}

The $w$ vortex with the magnetic $\pi$-rotation symmetry $P_3$ is described by the coefficients $C_{ab}(\rho)$ in Eq.~\eqref{eq:axivortex} as well as $v$ vortex.
However, the coefficients $C_{+0}$, $C_{0+}$, $C_{0-}$, and $C_{-0}$ are pure imaginary numbers.
Thus, the relative phase between the compensate A-phase and $\beta$-phase components and the bulk B-phase component is $\pi/2$, which is the difference of the OP from $v$ vortex.

\begin{figure}
\begin{center}
\includegraphics[width=8.5cm]{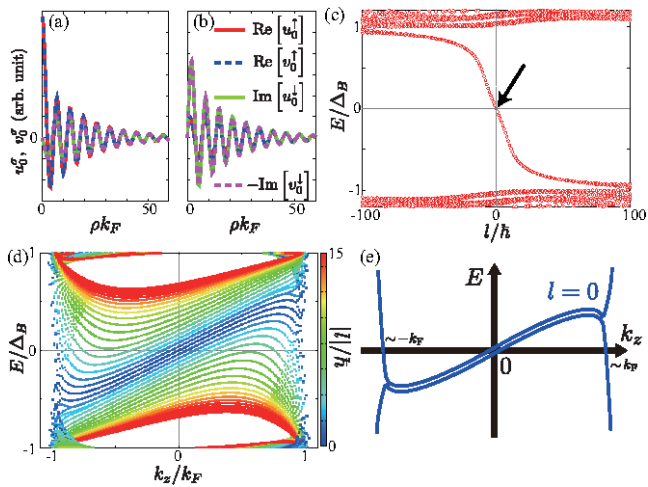}
\end{center}
\caption{\label{wvor}(Color online) Numerical results for $w$ vortex. 
(a) and (b) Degenerate wave functions $u^{\sigma}_0(\rho)$ and $v^{\sigma}_0(\rho)$ with $E_{\nu,k_z}=0$.
(c) Eigenvalues for quasiparticles with $k_z=0$. The degenerate eigenstates indicated by an arrow correspond to the two wave functions in (a) and (b).
(d) $k_z$-dispersion of the vortex bound state whose color denotes angular momentum $l$ of quasiparticles.
(e) Schematic $k_z$-dispersion for $l=0$ quasiparticles, which is precisely degenerate near $k_z=0$. 
}
\end{figure}

We calculate the vortex bound state for $w$ vortex by the BdG theory with the $v$ vortex OP shown in Fig.~\ref{vvor}(a) after changing the phases of $C_{+0}$, $C_{0+}$, $C_{0-}$, and $C_{-0}$ by $\pi/2$.
The obtained eigenvalues are shown in Figs.~\ref{wvor}(c) and \ref{wvor}(d).
In Fig.~\ref{wvor}(c), the eigenvalues for $k_z=0$ are classified into quantized orbital angular momentum $l$ along the vortex line.
The vortex bound state has degenerate two exact zero energy modes at $l=0$ indicated by an arrow.
The $k_z$-dispersion of the vortex bound state for various $l$'s is shown in Fig.~\ref{wvor}(d).
The quasiparticles with $l=0$ are degenerate in small $k_z$ and linearly cross the zero energy at $k_z=0$.
The $l=0$ branches cross the zero energy twice at $k_z=0$ and $|k_z|\sim k_{\rm F}$ schematically shown in Fig.~\ref{wvor}(e), which is consistent with the value of second Chern number discussed below.

The wave functions of the degenerate zero energy modes indicated by the arrow in Fig.~\ref{wvor}(c) are shown in Figs.~\ref{wvor}(a) and \ref{wvor}(b).
The quasiparticle wave functions for each mode satisfy ${\rm Re}\!\left(u_0^{\uparrow }\right)={\rm Re}\!\left(v_0^{\uparrow }\right)$ and ${\rm Im}\!\left(u_0^{\downarrow }\right)=-{\rm Im}\!\left(v_0^{\downarrow }\right)$, namely, $u_0^{\uparrow }=\left(v_0^{\uparrow }\right)^*$ and $u_0^{\downarrow }=\left(v_0^{\downarrow }\right)^*$.
Therefore, the degenerate zero energy modes are the Majorana zero modes.

The existence of the Majorana zero modes can be understood by the topological arguments in Appendix~\ref{app:clifford}.
The additional magnetic $\pi$-rotation symmetry $P_3$ in the
Altland-Zrinbauer symmetry class D gives a topological classification
$\mathbb{Z}\oplus\mathbb{Z}$ for the vortex bound state in three-dimensional space~\cite{shiozaki:2014}.

One of the topological numbers $\mathbb{Z}$ is the second Chern number.
For $w$ vortex, the second Chern number is found to be zero as well as $o$ vortex
because the BdG Hamiltonian for $w$ vortex
is identical to that for $o$ vortex at infinity from the vortex.
The vanishing second Chern number is consistent with the obtained
$k_z$-dispersion of the vortex bound states in Fig.~\ref{wvor}(d):
Some of vortex bound states cross the zero
energy in the $k_z$-direction, but the crossing always occurs twice, so
they can smoothly
merge into the bulk state without closing bulk gap.
Therefore, the zero energy states are topologically unstable, except
for the $l=0$ bound state.
The topological stability of the $l=0$ bound state is ensured by the other
$\mathbb{Z}$ topological invariant, namely the index ${\rm tr}\widehat{\Gamma }$.
Since the chiral symmetry $\widehat{\Gamma }=\widehat{\mathcal{C}}\widehat{\mathcal{P}}_3$ is also defined in the symmetric space $k_y=k_z=0$ and $\phi=0$ or $\pi$ for $w$ vortex, the one-dimensional winding numbers $w^{\phi=0}$ and $w^{\phi=\pi }$ are evaluated as the same manner in Sec.~\ref{sec:o_vortex}.
The difference of the winding numbers $(w^0-w^{\pi })/2=2$ provides the $\mathbb{Z}$ topological invariant and guarantees the existence of two zero energy states at $k_z=0$, which are indeed realized as the two $l=0$ zero modes in Fig.~\ref{wvor}.

\section{Summary}

We have calculated the bound state in $o$ vortex, $v$ vortex, and $w$ vortex by the full quantum BdG theory with the self-consistent OP obtained by using the quasiclassical theory.
Moreover, we have discussed symmetry protection of zero energy excitations in the vortex bound states with additional symmetry.
Our results are summarized in the following and in Table I.
Characteristic features of the bound states in $u$ vortex and $uvw$ vortex are also listed in Table I.

The most symmetric $o$ vortex has $P_1$, $P_2$, and $P_3$ symmetries in which induced components around the vortex core are fixed for $C_{++}$ and $C_{--}$ in real numbers, and $C_{+0}$, $C_{0+}$, $C_{-0}$, and $C_{0-}$ as zero.
The vortex bound state for the quasiparticles with angular momentum $l=0$ has spin degenerate Majorana zero modes at $k_z=0$.
The Majorana zero modes are protected by $P_3$ symmetry and characterized by a topological invariant $\mathbb{Z}$ which is the chiral index $(w^0-w^{\pi})/2=2$.

\begin{table*}
\begin{center}
\begin{minipage}{15cm}
\caption{Classified vortices by $P_1$, $P_2$, and $P_3$ symmetries which fix the element of induced components $C_{ab}(\rho)$.
For each vortex, we show the presence or absence of zero energy modes (ZEM) and Majorana zero modes, as well as topological invariant classifying the vortex bound state if it is present.
The candidates for the topological invariant are the chiral index $N\equiv(w^0-w^{\pi})/2=2$ and the second Chern number ${\rm Ch}_2=0$.}
\label{tab}
\end{minipage}
\begin{tabular*}{15cm}{@{\extracolsep{\fill}}ccccccc}
\hline \hline
\rule{0pt}{2.5ex} \multirow{2}{*}{vortex} & \multirow{2}{*}{symmetry} & \multirow{2}{*}{$C_{++}$, $C_{--}$} & $C_{+0}$, $C_{0+}$, & \multirow{2}{*}{ZEM} & \multirow{2}{*}{Majorana} & \multirow{2}{*}{top.~inv.} \\
\rule{0pt}{2.5ex} & & & $C_{-0}$, $C_{0-}$ \\ \hline
\rule{0pt}{2.5ex} $o$ vortex & $P_1$, $P_2$, $P_3$ & Real    & -- & $\checkmark$ & $\checkmark$ & $N$ \\
\rule{0pt}{2.5ex} axisym.~$v$ vortex & $P_2$               & Real    & Real & $\checkmark$ & -- & -- \\
\rule{0pt}{2.5ex} non-axisym.~$v$ vortex & $P_2$           & Real    & Real & -- & -- & -- \\
\rule{0pt}{2.5ex} $w$ vortex & $P_3$               & Real    & Imaginary & $\checkmark$ & $\checkmark$ & $N\oplus{\rm Ch}_2$ \\
\rule{0pt}{2.5ex} $uvw$ vortex & --                & Complex & Complex   & $\checkmark$ & --           & ${\rm Ch}_2$ \\
\rule{0pt}{2.5ex} $u$ vortex & $P_1$               & Complex & --        & $\checkmark$ & --           & --                           \\
\hline \hline
\end{tabular*}
\end{center}
\end{table*}

The $v$ vortex has $P_2$ symmetry in which all induced components around the vortex core are real numbers.
In the vortex bound state, the quasiparticles with $|l|\lesssim\xi_{\rm c}k_{\rm F}$ cross the zero energy twice at finite $\pm k_z$.
However, the zero energy modes are topologically trivial, that is, they are not protected any symmetry.
Then, low energy excitations in non-axisymmetric $v$ vortex has a gap because the deformation breaking axial symmetry lifts the zero energy modes in spite of keeping $P_2$ symmetry.
If we observe the excitation gap accompanying the vortex transition in the B-phase, that will be a strong evidence of that the realized vortex has only $P_2$ symmetry, namely, $v$ vortex.
Although the order of the gap is $\Delta/E_{\rm F}$, that becomes large for quasiparticles with $k_z\sim k_{\rm F}$ owing to the small effective Fermi energy $E_{\rm F}^{\rm 2D}=(\hbar^2/2m)(k_{\rm F}^2-k_z^2)$.
The excitation gap may have been observed as the difference of critical velocity between V1 and V2 vortices~\cite{pekola:1984b,pekola:1985}.

The $w$ vortex has $P_3$ symmetry.
The difference between $w$ vortex and $v$ vortex is only phases of induced components $C_{+0}$, $C_{0+}$, $C_{-0}$, and $C_{0-}$ concerning the OP.
However, the vortex bound state in superfluids belonging to the symmetry class D with additional symmetry $P_3$ is characterized by the topological invariants $\mathbb{Z}\oplus\mathbb{Z}$, the second Chern number and the chiral index.
For $w$ vortex, the second Chern number ${\rm Ch}_2=0$, but the chiral index $(w^0-w^{\pi})/2=2$ indicating difference between the numbers of the zero energy mode with opposite chirality.
Thus, the two-fold degenerate Majorana zero modes at the symmetric point $k_z=0$ and $l=0$ are protected by $P_3$ symmetry.

The $uvw$ vortex without any $P_1$, $P_2$, and $P_3$ symmetries has many zero energy modes, as calculated by Silaev~\cite{silaev:2009}.
However, the zero energy modes are topologically unstable because the vortex bound state in superfluids belonging to the class D without additional symmetry is characterized by the second Chern number, which is zero for the vortex bound state in the superfluid $^3$He B-phase.
The $u$ vortex with $P_1$ symmetry is achieved by adding imaginary parts to real $C_{++}$ and $C_{--}$ components in $o$ vortex.
The imaginary parts lift the Majorana zero modes at the symmetric point $k_z=0$ and $l=0$ by breaking $P_3$ symmetry.
Other accidental zero modes will be left but they are not characterized by the second Chern number which generally vanishes under $P_1$ symmetry.

In conclusion, Majorana zero modes protected by $P_3$ symmetry are bound in $o$ vortex and $w$ vortex which are not stable in the bulk B-phase.
The $v$ vortex is more stable than the $o$ vortex owing to the condensation energy by A-phase and $\beta$-phase components which compensates the vortex core.
Then, when we confine the B-phase in a thin slab with hight along the vortex line to suppress the $\beta$-phase component and simultaneously apply a magnetic field along the vortex line to suppress the A-phase component, $o$ vortex with Majorana zero modes will be achieved.
Note that the confinement and the magnetic field do not break any $P_1$, $P_2$, and $P_3$ symmetries.

%

\section*{Acknowledgments}
We thank M.~Ichioka, T.~Mizushima, T.~Morimoto, A.~Furusaki, and S.~Kobayashi for helpful discussion.
A part of the numerical calculations was performed by using the RIKEN Integrated Cluster of Clusters (RICC) and ICE8200EX in NIMS.
This work was supported by KAKENHI (Nos. 24840048, 21340103, 2200247703, 25287085, and 22103005) and WPI Initiative on Material nanoarchtectonics, MEXT, Japan.
K.S.~is supported by a JSPS Fellowship for Young Scientists.

\appendix
\section{Linear $k_z$-dispersion of $o$ vortex bound state}
\label{app:ovortex}

We show the linear $k_z$-dispersion of the vortex bound state in $o$ vortex on the basis of perturbation theory~\cite{kawakami:2011}. 
The gap function in Eq.~\eqref{fourier} for $o$ vortex is described by
\begin{multline}
\hat{\Delta}_{k_z}(\bm{\rho}_1, \bm{\rho}_2)\\
=\int \frac{d\bm{k}_{\mathrm{2D}}}{(2\pi)^2}e^{i\bm{k}_{\mathrm{2D}}\cdot\bm{\rho}'}\frac{\Delta_0(\rho)e^{i\phi}}{k_{\rm F}}
\begin{pmatrix}
-\!k_x\!+\!ik_y &k_z \\
k_z & k_x\!+\!ik_y
\end{pmatrix},
\end{multline}
which are simplified as $C_{+-}=C_{00}=C_{-+}=\Delta_0({\rho })e^{i\phi }$, $C_{++}=C_{--}=0$, and $\Gamma(k)=1$.
For $k_z\!=\!0$, the BdG Hamiltonian in Eq.~\eqref{2DBdG} is separated into $\uparrow$-spin and $\downarrow$-spin sectors and each sector has a Majorana zero mode. The wave functions of the Majorana zero modes are described by
\begin{eqnarray}
\vec{u}_0^{\uparrow}(\bm{\rho}) &=& 
\left(u_0^\uparrow(\rho),\ 0,\
\left[u_0^\uparrow(\rho)\right]^*,\ 0\right)^T, \label{zesu} \\
\vec{u}_0^{\downarrow}(\bm{\rho}) &=& \left(0,\ u_0^\downarrow(\rho) e^{i\phi},\ 0,\ \left[u_0^\downarrow(\rho)\right]^* e^{-i\phi}\right)^T\label{zesd},
\end{eqnarray}
where $u_0^{\uparrow}(\rho)=\mathcal{N}_{\uparrow}J_0(k_{\rm F}\rho)\exp(-\rho/\xi\!-\!i\pi/4)$ and $u_0^{\downarrow}(\rho)=\mathcal{N}_{\downarrow}J_1(k_{\rm F}\rho)\exp(-\rho/\xi+i\pi/4)$ with normalization factors $\mathcal{N}_\sigma$ and the $n$-th order Bessel function $J_n$~\cite{gurarie:2007,mizushima:2010}.
The wave function of the Majorana zero mode in each spin sector differs in the phase winding and the order of the Bessel function since the Hamiltonian for the $\uparrow$-spin ($\downarrow$-spin) sector corresponds to the Hamiltonian for the chiral $p$-wave superfluid with the chirality antiparallel (parallel) to the vorticity.

Here, we consider the $k_z$-dispersion with small $k_z\ll k_{\rm F}$ by perturbation theory.
The perturbation Hamiltonian is described by
\begin{eqnarray}
\widehat{H}_{\mathrm{p}}({\bm\rho})\!=\!\frac{k_z}{k_{\rm F}}{\Delta}_0(\rho)
\begin{pmatrix}
0 &0 &0 &e^{i\phi}\\
0 &0 &e^{i\phi} &0 \\
0 &e^{-i\phi} & 0 &0 \\
e^{-i\phi} &0 & 0 &0 
\end{pmatrix},
\end{eqnarray}
where projections of $\widehat{H}_{\mathrm{p}}$ onto $\vec{u}_0^{\uparrow }$ and $\vec{u}_0^{\downarrow }$ give
\begin{multline}
\left(\begin{array}{cc}
\left[\vec{u}_0^{\uparrow}({\bm\rho})\right]^{\dag}\widehat{H}_{\mathrm{p}}({\bm\rho})\vec{u}_0^{\uparrow}({\bm\rho})&
\left[\vec{u}_0^{\uparrow}({\bm\rho})\right]^{\dag}\widehat{H}_{\mathrm{p}}({\bm\rho})\vec{u}_0^{\downarrow}({\bm\rho})\\
\left[\vec{u}_0^{\downarrow}({\bm\rho})\right]^{\dag}\widehat{H}_{\mathrm{p}}({\bm\rho})\vec{u}_0^{\uparrow}({\bm\rho})&
\left[\vec{u}_0^{\downarrow}({\bm\rho})\right]^{\dag}\widehat{H}_{\mathrm{p}}({\bm\rho})\vec{u}_0^{\downarrow}({\bm\rho})
\end{array}\right)\\
=2\mathcal{N}_\uparrow\mathcal{N}_\downarrow \frac{k_z}{k_{\rm F}}\Delta_0(\rho) J_0(k_{\rm F}\rho)J_1(k_{\rm F}\rho) e^{-2\rho/\xi }\hat{\sigma}_x. \label{Hp}
\end{multline}
Then, the $k_z$-dispersion,
\begin{align}
E_{\pm} = \pm 4\pi\mathcal{N}_\uparrow\mathcal{N}_\downarrow \frac{k_z}{k_{\rm F}}\int d\rho\Delta_0(\rho)J_0(k_{\rm F}\rho)J_1(k_{\rm F}\rho)e^{-2\rho/\xi },
\end{align}
is linear to $k_z$.
The 0-th order perturbed wave functions are
\begin{eqnarray}
\vec{u}_{0}^{\pm}(\bm{\rho})=\frac{1}{\sqrt{2}}[\vec{u}_{0}^{\uparrow}(\bm{\rho})\pm\vec{u}_{0}^{\downarrow}(\bm{\rho})].
\end{eqnarray}
These wave functions $\vec{u}_0^{\pm}(\bm{\rho})$ also indicate self-conjugate Majorana quasiparticles.

\section{Topological classification of vortex bound states by Clifford algebras}
\label{app:clifford}

In this appendix, we discuss topological classification of vortex bound
states in the B-phase.
Consider a vortex along the $z$-axis, and a circle surrounding the
vortex, which is parametrized by the angle $\phi$ evaluated from the $x$-axis.
The semi-classical BdG Hamiltonian on the circle is given by
%
\begin{align}
\widehat{H}_{\rm BdG}({\bm k},\phi)=
\begin{pmatrix}
\hat{h}({\bm k}) & \hat{\Delta }({\bm k},\phi) \\
\hat{\Delta }^{\dagger }({\bm k},\phi) & -\hat{h}^{\rm T}(-{\bm k})
\end{pmatrix},
\end{align}
where $\hat{h}({\bm k})=(\hbar^2/2m)({\bm k}^2-k_{\rm F}^2)\hat{1}$ is the
Hamiltonian in the normal state, and $\hat{\Delta }$ is OP
which approaches
\begin{eqnarray}
\hat{\Delta}({\bm k}, \phi)=\frac{\Delta_{\rm B}}{k_{\rm F}}
\left(
\begin{array}{cc}
-k_x+ik_y & k_z\\
k_z & k_x+ik_y
\end{array}
\right)e^{i\phi}, 
\end{eqnarray}
far away from the vortex core.
The BdG Hamiltonian has particle-hole symmetry defined by
$\widehat{\mathcal{C}}\widehat{H}_{\rm BdG}({\bm
k},\phi)\widehat{\mathcal{C}}^{-1}=-\widehat{H}_{\rm BdG}(-{\bm
k},\phi)$ with $\widehat{\mathcal{C}}=\widehat{\tau }_x\mathcal{K}$ and
$\widehat{\mathcal{C}}^2=\widehat{1}$, where $\mathcal{K}$ is the
complex conjugation operator.

The axisymmetric vortices in the $^3$He B-phase may have three types of
discrete symmetries: $P_1$, $P_2$, and
$P_3$~\cite{salomaa:1987}.
The inversion symmetry $P_1$ implies 
$\widehat{H}_{\rm BdG}({\bm
k},\phi)=\widehat{H}_{\rm BdG}(-{\bm k},\phi+\pi)$.
$P_2$ is magnetic reflection symmetry that obtained as a combination of
time-reversal and mirror reflection with respect to a plane including
the vortex line: 
If we take the $xz$-plane as the reflection plane, $P_2$ reads
$\widehat{H}_{\rm BdG}({\bm
k},\phi)=\widehat{\mathcal{P}}_2\widehat{H}_{\rm
BdG}(-k_x,k_y,-k_z,-\phi)\widehat{\mathcal{P}}_2^{-1}$, where
$\widehat{\mathcal{P}}_2=\hat{\sigma }_y\widehat{\tau
}_z\widehat{\mathcal{T}}=i\widehat{\tau }_z\mathcal{K}$ with the
time-reversal operator $\widehat{\mathcal{T}}=i\hat{\sigma
}_y\mathcal{K}$. 
$P_3$ is magnetic $\pi$-rotation symmetry around an axis perpendicular
to the vortex line, say, the $x$-axis.
$P_3$
means $\widehat{H}_{\rm
BdG}({\bm
k},\phi)=\widehat{\mathcal{P}}_3\widehat{H}_{\rm
BdG}(-k_x,k_y,k_z,-\phi)\widehat{\mathcal{P}}_3^{-1}$, where
$\widehat{\mathcal{P}}_3=i\hat{\sigma }_z\widehat{\tau
}_z\mathcal{K}$. 
Among these discrete symmetries, we respect $P_2$ and/or $P_3$ in the following,
since $P_1$ does not provide symmetry protected defect zero modes in general.
Below we identify all topological invariants relevant to
existing vortex zero modes in the
B-phase.

First note that the BdG Hamiltonian far way from the vortex core can be
written in terms of the gamma matrices $\gamma_{\mu}=(-\sigma_z\tau_x, -\tau_y, \sigma_x\tau_x,-\tau_z)$ as
\begin{align}
\widehat{H}_{\rm BdG}({\bm k}, \phi)=e^{i(\phi/2)
 \gamma_4}d_{\mu}\gamma_{\mu}e^{-i(\phi/2)\gamma_4}
-\frac{\hbar^2{\bm k}^2}{2m}\gamma_4,
\label{eq:BdGawaycore}
\end{align} 
with $d_{\mu}=(\Delta_{\rm B}k_x/k_{\rm F},\Delta_{\rm B}k_y/k_{\rm F},\Delta_{\rm B}k_z/k_{\rm F},
\hbar^2k_{\rm F}^2/2m)$.
The gamma matrices $\gamma_{\mu}$ obey the Clifford algebra,
$\{\gamma_\mu, \gamma_{\nu}\}=2\delta_{\mu\nu}$.
Whereas the BdG Hamiltonian near the core is different from 
Eq.(\ref{eq:BdGawaycore}), 
it is smoothly interpolated from this, 
with keeping symmetry of the vortex. 
Therefore, the topological structure of a vortex can be evaluated from
the asymptotic Hamiltonian Eq.(\ref{eq:BdGawaycore}), subject to a set
of symmetries of the vortex.

To clarify the possible topological structure, we
furthermore deform Eq.~(\ref{eq:BdGawaycore})
into the form of a Dirac Hamiltonian. 
For this purpose, we may add the following topologically trivial system
with the same discrete symmetries,
\begin{align}
\widehat{H}'_{\rm BdG}({\bm k}, \phi)=-e^{-i(\phi/2)
 \gamma_4}d_{\mu}\gamma_{\mu}e^{i(\phi/2)\gamma_4}-\frac{\hbar^2{\bm
 k}^2}{2m}\gamma_4.
\end{align} 
In comparison with the original BdG Hamiltonian, $\widehat{H}'_{\rm
BdG}({\bm k}, \phi)$ has a negative
chemical potential $-\hbar^2k_{\rm F}^2/2m$, and thus it
is deformable to a topologically trivial insulator without gap closing, by
taking the limit $\Delta_{\rm B}\rightarrow 0$. 
This means that $\widehat{H}'_{\rm BdG}({\bm k}, \phi)$ is
topologically trivial.
Adding the topological trivial band to the original one, we have the
extended BdG Hamiltonian $\widehat{H}_{\rm eBdG}({\bm k}, \phi)$ given by
the direct product of $\widehat{H}_{\rm BdG}({\bm k}, \phi)$ and
$\widehat{H}'_{\rm BdG}({\bm k}, \phi)$, 
\begin{eqnarray}
\widehat{H}_{\rm eBdG}({\bm k}, \phi)
=\left(
\begin{array}{cc}
\widehat{H}_{\rm BdG}({\bm k}, \phi) & 0\\
0 & \widehat{H}'_{\rm BdG}({\bm k}, \phi) 
\end{array}
\right), 
\end{eqnarray}
which is again written in terms of the gamma
matrices as
\begin{align}
\widehat{H}_{\rm eBdG}({\bm k}, \phi)= e^{i(\phi/2)
 \Gamma_4}d_{\mu}\Gamma_{\mu}e^{-i(\phi/2)\Gamma_4}-\frac{\hbar^2{\bm
 k}^2}{2m}\gamma_4,
\end{align}
where the new gamma matrices $\Gamma_\mu$ are given by
$\Gamma_{\mu}=\gamma_{\mu}\mu_z$ with the Pauli
matrix $\mu_i$ in the band space.
The particle-hole and other discrete symmetries imply that
\begin{eqnarray}
&&[\mathcal{C},\Gamma_1]=[\mathcal{C},\Gamma_2]=[\mathcal{C},\Gamma_3]
=\{\mathcal{C},\Gamma_4\}=0,
\label{eq:c1}\\
&&\{\mathcal{P}_2,\Gamma_1\}=[\mathcal{P}_2,\Gamma_2]=\{\mathcal{P}_2,\Gamma_3\}
=[\mathcal{P}_2,\Gamma_4]=0,
\label{eq:c2}
\\
&&\{\mathcal{P}_3,\Gamma_1\}=[\mathcal{P}_3,\Gamma_2]=[\mathcal{P}_3,\Gamma_3]
=[\mathcal{P}_3,\Gamma_4]=0,
\label{eq:c3}
\end{eqnarray}
with
\begin{eqnarray}
&&[\mathcal{C},\mathcal{P}_2]=[\mathcal{C},\mathcal{P}_3]=[\mathcal{P}_2, \mathcal{P}_3]=0,
\label{eq:c5}
\end{eqnarray}
($\gamma_4$ obeys the same (anti-)commutation
relations as $\Gamma_4$.)
The extended BdG Hamiltonian is stable-equivalent to the original one in
the sense of the K-theory.


For the extended BdG Hamiltonian, we can introduce $\Gamma_5=\mu_y$ that
satisfies 
\begin{eqnarray}
&&\Gamma_5^2=1, \quad
\{\Gamma_5, \Gamma_{\mu}\}=0,
\\
&&\{{\cal C},\Gamma_5\}=\{{\cal P}_2,\Gamma_5\}=\{{\cal P}_3, \Gamma_5\}=0.
\end{eqnarray}
Using $\Gamma_5$, we perform the deformation of the
Hamiltonian that preserves all the discrete symmetries (as well as the
particle-hole symmetry) except for $P_1$: 
\begin{eqnarray}
&&\widehat{H}_\alpha({\bm k}, \phi)= U_{\alpha}(\phi)d_{\mu}\Gamma_{\mu}
U_{\alpha}^{-1}(\phi)
-\frac{\hbar^2{\bm
 k}^2}{2m}\gamma_4,
\end{eqnarray}
where $U_{\alpha}(\phi)$ is given by
\begin{eqnarray}
U_{\alpha}(\phi)=e^{i(\phi/2)[\cos\alpha \Gamma_4+i\sin\alpha
 \Gamma_4\Gamma_5]}.
\end{eqnarray}
Through this equation, the extended BdG Hamiltonian at $\alpha=0$ is
smoothly deformed into the following Dirac Hamiltonian at
$\alpha=\pi/2$,
\begin{eqnarray}
\widehat{H}_{\rm D}({\bm k}, \phi)=k_x\Gamma_1+k_y\Gamma_2+k_z\Gamma_3
+\cos\phi\Gamma_4+\sin\phi\Gamma_5,
\label{eq:Dirac}
\nonumber\\ 
\end{eqnarray}
where we have placed $\Delta_{\rm B}/k_{\rm F}=1$ and $d_4=1$ and
omitted the regularization term $-(\hbar^2{\bm k}^2/2m)\gamma_4$ for simplicity. 
The Dirac Hamiltonian has the same topological properties as the
original BdG Hamiltonian.
To elucidate the topological structure of the Dirac Hamiltonian, 
we consider a family of Dirac Hamiltonians which have the same form and
the same symmetries as Eq.~(\ref{eq:Dirac}).
Even for  these Dirac Hamiltonians, 
 $\Gamma_{\mu}$, $\Gamma_5$, ${\cal C}$, ${\cal P}_2$, and ${\cal P}_3$
 should satisfy the same commutation or anti-commutation
 relations as Eqs.~(\ref{eq:c1})-(\ref{eq:c5}), but their matrix
 representation are not specified anymore. 
For the family of Dirac Hamiltonians,
$\Gamma_4$ transforms like a mass term
under the symmetries, so we can clarify the topological structure
by counting topologically distinct $\Gamma_4$ matrices consistent with the
symmetries~\cite{kitaev:2009, morimoto:2013}.

To count topologically distinct $\Gamma_4$ matrices, we use the fact that the
symmetry operators and gamma matrices form a real Clifford algebra
$Cl_{p,q}$ that has $p+q$ generators $\{e_1,\dots, e_{p}; e_{p+1},\dots,
e_{p+q}\}$ satisfying
\begin{eqnarray}
&&\{e_i,e_j\}=0, \,(i\neq j),
\nonumber\\
&&e_i^2=\left\{
\begin{array}{cl}
-1, & 1\le i \le p\\
+1, & p+1 \le i \le p+q
\end{array}
\right. .
\end{eqnarray} 
For instance, consider $uvw$ vortex that only has the particle-hole
symmetry.
The particle-hole symmetry ${\cal C}$ and the gamma
matrices other than $\Gamma_4$ form $Cl_{3,3}$ as
\begin{align}
\{
\Gamma_1\mathcal{J},\Gamma_2\mathcal{J},\Gamma_3\mathcal{J};\mathcal{C},\mathcal{C}\mathcal{J},\Gamma_5\}, 
\end{align}
with
\begin{align}
 (\Gamma_i\mathcal{J})^2=-1,\quad
\mathcal{C}^2=1,\quad (\mathcal{C}\mathcal{J})^2=1,
\quad \Gamma_5^2=1.
\end{align}
Here we have introduced ${\cal J}$ representing the imaginary unit
``$i$'' so that we can treat complex structure originated from the anti-unitary
operator ${\cal C}$. 
If we take into account $\Gamma_4$ as well, the Clifford algebra is
extended to $Cl_{3,4}$.
\begin{align}
\{
\Gamma_1\mathcal{J},\Gamma_2\mathcal{J},\Gamma_3\mathcal{J};\mathcal{C},\mathcal{C}\mathcal{J},\Gamma_5,
 \Gamma_4 \}. 
\end{align}
 Therefore, having a $\Gamma_4$ matrix
consistent with the particle-hole symmetry provides an
 extension of the Clifford algebra from $Cl_{3,3}$ to $Cl_{3,4}$, and
 vice versa.
A set of the latter extensions defines the classifying
space ${\cal R}_0$, so topologically distinct $\Gamma_4$ matrices can be
counted as the number of the disconnected parts of the classifying
space, i.e. $\pi_0({\cal R}_0)=\mathbb{Z}$.
Correspondingly, we can introduce the second Chern
 number defined in Eq.~(\ref{eq:secondCh})~\cite{teo:2010,qi:2008}.
For a vortex in $^3$He-B phase, however, the second Chern number becomes
zero since its asymptotic Hamiltonian Eq.~(\ref{eq:BdGawaycore}) has an
additional inversion symmetry.
(See also discussions in Sec.~\ref{sec:o_vortex}.)

For $o$ vortex, the corresponding Dirac Hamiltonian should be
subject to additional $P_2$ and $P_3$ symmetries. 
%
%
%
In the presence of $P_2$ and $P_3$, symmetry operators and the gamma
matrices other than $\Gamma_4$ form $Cl_{4,4}$ as
\begin{eqnarray}
\{\Gamma_1\mathcal{J},\Gamma_2\mathcal{J},\Gamma_3\mathcal{J},\mathcal{C}\Gamma_2\Gamma_5\mathcal{P}_2;
 \mathcal{C},\mathcal{C}\mathcal{J},
\Gamma_5,\mathcal{J}\Gamma_3\mathcal{P}_2\mathcal{P}_3\},
\nonumber\\
\end{eqnarray}
then a $\Gamma_4$ matrix extends this into $Cl_{4,5}$ as 
\begin{eqnarray}
\{\Gamma_1\mathcal{J},\Gamma_2\mathcal{J},\Gamma_3\mathcal{J},\mathcal{C}\Gamma_2\Gamma_5\mathcal{P}_2;
 \mathcal{C},\mathcal{C}\mathcal{J},
\Gamma_5,\mathcal{J}\Gamma_3\mathcal{P}_2\mathcal{P}_3, \Gamma_4\}.
\nonumber\\
\end{eqnarray}
As the classifying space for the extension $Cl_{4,4}\to Cl_{4,5}$ is  $R_0$, 
the BdG Hamiltonian with $o$ vortex
is classified as $\pi_0(R_0)=\mathbb{Z}$. 
The corresponding topological number is $(w^0-w^{\pi})/2$ 
defined in Eq.~(\ref{eq:1dwinding})~\cite{shiozaki:2014}.

For $v$ vortex, the Dirac Hamiltonian has additional $P_2$ symmetry.
Possible $\Gamma_4$ matrices can be identified as the extension of
Clifford algebra $Cl_{4,3}\to Cl_{4,4}$:
%
%
%
\begin{multline}
\{
\Gamma_1\mathcal{J},\Gamma_2\mathcal{J},\Gamma_3\mathcal{J},
\mathcal{C}\Gamma_2\Gamma_5\mathcal{P}_2;
\mathcal{C},\mathcal{C}\mathcal{J},\Gamma_5
\}
\\
\to
\{
\Gamma_1\mathcal{J},\Gamma_2\mathcal{J},\Gamma_3\mathcal{J},\mathcal{C}\Gamma_2\Gamma_5\mathcal{P}_2;
\mathcal{C},\mathcal{C}\mathcal{J},\Gamma_5, \Gamma_4
\}.
\end{multline}
%
The classifying space for the extension $Cl_{4,3}\to Cl_{4,4}$ is $R_{-1}\simeq R_7$.
Since $\pi_0(R_7)=0$, $v$ vortex does not support a zero mode protected
by $P_2$ symmetry.

Finally, consider $w$ vortex. 
In this case, the additional symmetry is $P_3$.
Topologically distinct $\Gamma_4$ matrices can be identified by using the
extension
%
%
%
\begin{multline}
\{
\Gamma_1\mathcal{J},\Gamma_2\mathcal{J},\Gamma_z\mathcal{J};
\mathcal{C},\mathcal{C}\mathcal{J}, \Gamma_5\}\otimes\{\mathcal{P}'_3\}\\
\to
\{
\Gamma_1\mathcal{J},\Gamma_2\mathcal{J},\Gamma_z\mathcal{J};
\mathcal{C},\mathcal{C}\mathcal{J}, \Gamma_5, \Gamma_4\}\otimes\{\mathcal{P}_3'\},
\end{multline}
with $\mathcal{P}_3'=\mathcal{C}\mathcal{J}\Gamma_2\Gamma_3\Gamma_5\mathcal{P}_3$.
This gives $Cl_{3,3}\otimes Cl_{0,1}\to Cl_{3,4}\otimes Cl_{0,1}$, which
classifying space is given by $R_0\times R_0$.
Therefore the topological classification of the BdG Hamiltonian with
$w$ vortex is $\pi_0(R_0\times R_0)=\mathbb{Z}\oplus\mathbb{Z}$.
The corresponding topological invariants are the second Chern
number in Eq.~(\ref{eq:secondCh})
and the
one-dimensional winding number 
defined in Eq.~(\ref{eq:1dwinding}).

\section{Phase shift across a vortex core}
\label{app:shift}

In this appendix, we make a detailed explanation for the excitation spectrum shown in Fig.~\ref{vvor}(e) in terms of the OP structure. 
When a vortex core is compensated by superfluid components, the vortex bound state for quasiparticles across the vortex core can be regarded as the Andreev bound state on the junction with the compensated superfluid between the bulk B-phases.
If the OPs of them are simultaneously diagonalizable on a quasiparticle path, we can estimate the bound state energy from the phase shift on the vortex core.

The explicit OPs in the bulk B-phase and for the induced superfluid in the vicinity of an axisymmetric vortex are described by
\begin{align}\label{eq:bulkb}
\hat{\Delta}_\mathrm{B}(\bar{\bm{k}},\bm{\rho})\!=\!
\Delta_\mathrm{B}(\rho)e^{i\phi}\left(\begin{array}{cc}
-\sin\theta_k e^{-i\phi_k} & \cos\theta_k \\
\cos\theta_k & \sin\theta_ke^{i\phi_k}
\end{array}\right),
\end{align}
and
\begin{widetext}\begin{align}\label{eq:core0}
	\hat{\Delta}_{\mathrm{core}}(\bar{\bm{k}},\bm{\rho})=
	\begin{pmatrix}
		C_{+0}(\rho)\cos\theta_k\!+\!C_{++}(\rho)\sin\theta_k e^{i(\phi_k-\phi)} &
		C_{0+}(\rho)\sin\theta_k e^{i\phi_k}\!-\!C_{0-}(\rho)\sin\theta_k e^{i(2\phi-\phi_k)}\\
		C_{0+}(\rho)\sin\theta_k e^{i\phi_k}\!-\!C_{0-}(\rho)\sin\theta_k e^{i(2\phi-\phi_k)}& 
		C_{-0}(\rho)\cos\theta_k e^{2i\phi} - C_{--}(\rho)\sin\theta_k e^{i(3\phi_k-\phi)}\\
	\end{pmatrix},
\end{align}\end{widetext}
where $\phi_k$ and $\theta_k$ denote the azimuthal and polar angles of the quasiparticle momenta, respectively, and $\phi$ is an azimuthal angle of the real space coordinate.
Since slightly induced components $C_{++}$ and $C_{--}$ do not influence phase shifts of the OP, we disregard them in the following discussion.

\begin{figure}
\begin{center}
\includegraphics[width=7.5cm]{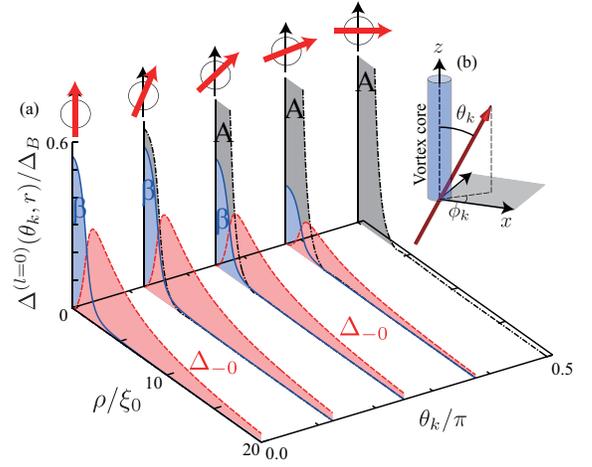}
\end{center}
\caption{\label{qpvvor}(Color online) (a) OP components $\Delta_{\rm A}$ (dash-dotted line), $\Delta_\beta$ (solid line), and $\Delta_{-0}$ (dashed line) in Eq.~\eqref{eq:core} as a function of the quasiparticle momentum $\theta_k$ and distance from a vortex core $\rho$. (b) Thick arrow denotes a quasiparticle path defined by $\theta_k$ and $\phi_k$.}
\end{figure}

Let us focus our attention on the quasiparticle path across the vortex core shown in Fig.~\ref{qpvvor}(b). The azimuthal angles of the quasiparticle momentum and the real space are $\phi\!=\!\phi_k$ or $\phi\!=\!\phi_k+\pi$, where the quasiparticle has the angular momentum $l\!=\!0$. For the quasiparticles with $l=0$, the OP of the vortex core state in Eq.~\eqref{eq:core0} is simplified as
\begin{multline}\label{eq:core}
\hat{\Delta}^{l\!=\!0}_{\mathrm{core}}(\bar{\bm{k}},\rho)
=\Delta_{\mathrm{A}}(\theta_k, \rho)e^{i\phi_k}\hat{\sigma}_x
+\Delta_{\mathrm{\beta}}(\theta_k,\rho)\frac{\hat{1}\!+\!\hat{\sigma}_z}{2}\\
+\Delta_{\mathrm{-0}}(\theta_k, \rho)e^{i\phi_k}e^{-i\phi_k\hat{\sigma}_z},
\end{multline}
where $\Delta_{\rm A}(\theta_k, \rho)=(C_{0+}-C_{0-})\sin\theta_k$, $\Delta_{\beta }(\theta_k, \rho)=(C_{+0}-C_{-0})\cos\theta_k$, and $\Delta_{-0}(\theta_k, \rho)=C_{-0}\cos\theta_k$.
Figure~\ref{qpvvor}(a) shows the $\theta_{k}$- and $\rho$-dependence of the OP components $\Delta_{\rm A}(\theta_k,\rho)$, $\Delta_{\beta }(\theta_k,\rho)$, and $\Delta_{-0}(\theta_k,\rho)$, which are obtained from our numerical results in Fig.~\ref{vvor}(a). 

As shown in Fig.~\ref{qpvvor}(a), the A-phase component $\Delta_A$ dominates at the vortex core for $\theta_k\!=\!\pi/2$. The quasiparticles with momentum perpendicular to the vortex line can be regarded as quasiparticles across the junction with the A-phase between the bulk B-phase. The OPs of the bulk B-phase in Eq.~\eqref{eq:bulkb} and the A-phase described by the $\Delta_A$ term in Eq.~\eqref{eq:core} are simultaneously diagonalizable by using the unitary matrix 
\begin{align}\label{eq:u1}
\hat{U}\!=\!\exp\left[-i(\pi/4)(\bar{\bm{k}}\cdot\hat{\bm{\sigma}})\right].
\end{align}
Then, they are transformed as
\begin{multline}
\hat{U}\hat{\Delta}_{\mathrm{B}}(\phi_k,\theta_k=\pi/2,\rho,\phi=\phi_k,\phi_k+\pi)\hat{U}^T\\
= \pm\Delta_{\rm B}(\rho)e^{i\phi_k}e^{-i\phi_k\hat{\sigma}_z}\left(\begin{array}{cc}
-1 &0 \\
0 & 1 
\end{array}\right),
\label{eq:Btrans}
\end{multline}
where $+$ $(-)$ sign is taken for $\phi=\phi_k$ $(\phi=\phi_k+\pi)$,
and
\begin{multline}
\hat{U}\Delta_{\mathrm{A}}(\theta_k=\pi/2,\rho)e^{i\phi_k}\hat{\sigma_x}\hat{U}^T \\
= -[C_{0+}(\rho)-C_{0-}(\rho)]e^{i\phi_k}e^{-i\phi_k\hat{\sigma}_z}\left(\begin{array}{cc}
i &0 \\
0 &i 
\end{array}\right).\label{eq:Atrans}
\end{multline}

For $v$ vortex with real $C_{0+}$ and $C_{0-}$, comparison of Eqs.~\eqref{eq:Btrans} and \eqref{eq:Atrans} shows that the quasiparticles feel the phase shift $\varphi=\pm\pi/2$ for each spin sector. The eigenenergy of the Andreev bound states on a junction with phase shift $\varphi$ is $E\propto\pm|\cos(\varphi/2)|$~\cite{furusaki:1999,kashiwaya:2000}. Therefore, quasiparticles with momentum $\theta_k=\pi/2$ and angular momentum $l\!=\!0$ have an energy gap in the vortex bound state. This gap corresponds to the energy gap at $k_z=0$ and $l\!=\!0$ in Figs.~\ref{vvor}(d) and \ref{vvor}(e). 

On the other hand, Fig.~\ref{qpvvor}(a) shows that the $\Delta_{-0}$ term dominates around the vortex core for the quasiparticle with momentum $\theta_k=0$.
Note that $\Delta_{-0}$ and $\Delta_{\beta }$ induced on the vortex core have the same phase factor.
We can simultaneously diagonalize the OPs of the bulk B-phase in Eq.~\eqref{eq:bulkb} and the $\Delta_{-0}$ term in Eq.~\eqref{eq:core} by using the unitary matrix 
\begin{align}\label{eq:u2}
\hat{U}=\exp\left[-\frac{i}{2}\left(\frac{\pi}{2}\!-\!\theta_k\right)(\bar{\bm e}_{\phi_k}\cdot\bm{\sigma})\right],
\end{align}
where $\bar{\bm e}_{\phi_k}=-\bar{\bm x}\sin\phi_k+\bar{\bm y}\cos\phi_k$.
After the unitary transformation at $\theta_k=0$, the OP of the $\Delta_{-0}$ term remains unchanged and that of the bulk B-phase transforms into the same form in Eq.~\eqref{eq:Btrans}. Therefore, quasiparticles with momentum $\bm{k}$ almost parallel to the vortex line feel the phase shift $\varphi=\pi$. This $\pi$-phase shift results in the $l\!=\!0$ excitation spectrum approaching the zero energy at $k_z=\pm k_{\rm F}$ as shown in Fig.~\ref{vvor}(e). 

In this sense, the $\Delta_{-0}$ term plays a crucial role in the excitation spectrum. If we set only $C_{+0}$ and $C_{0+}$ components to be non-zero near the vortex core as Ref.~\cite{silaev:2009}, we can diagonalize the OP of the bulk B-phase and compensated components in the vortex core simultaneously by using the unitary matrix in Eq.~\eqref{eq:u2} with $\theta_k\!=\!\arccos[\pm\sqrt{2C_{0+}/(2C_{0+}-C_{+0})}]$, where $C_{0+}$ and $C_{+0}$ have opposite signs. Since quasiparticles with the momenta $k_z=\pm k_{\rm F}|\cos\theta_k|$ feel the $\pi$-phase shift, the $l=0$ excitation spectrum should cross the zero energy in $0<|k_z|<k_{\rm F}$. However, we can obtain $l=0$ excitation spectrum crossing the zero energy at $k_z=\pm k_{\rm F}$ because we consider not only $C_{+0}$ and $C_{0+}$ but also the other induced components. 

The phase shift also explains the existence of the degenerate Majorana zero modes in $w$ vortex. For $w$ vortex, since $C_{0+}$ and $C_{0-}$ in Eq.~\eqref{eq:Atrans} are pure imaginaries, quasiparticles with $k_z=0$ and $l=0$ feel the $\pi$-phase shift, which is confirmed by comparison of Eqs.~\eqref{eq:Btrans} and \eqref{eq:Atrans}. The zero energy bound states at the particle-hole symmetric point owing to the $\pi$-phase shift are exactly degenerate Majorana zero modes.

\bibliographystyle{apsrev4-1}
\bibliography{references}

\begin{thebibliography}{47}%
\makeatletter
\providecommand \@ifxundefined [1]{%
 \@ifx{#1\undefined}
}%
\providecommand \@ifnum [1]{%
 \ifnum #1\expandafter \@firstoftwo
 \else \expandafter \@secondoftwo
 \fi
}%
\providecommand \@ifx [1]{%
 \ifx #1\expandafter \@firstoftwo
 \else \expandafter \@secondoftwo
 \fi
}%
\providecommand \natexlab [1]{#1}%
\providecommand \enquote  [1]{``#1''}%
\providecommand \bibnamefont  [1]{#1}%
\providecommand \bibfnamefont [1]{#1}%
\providecommand \citenamefont [1]{#1}%
\providecommand \href@noop [0]{\@secondoftwo}%
\providecommand \href [0]{\begingroup \@sanitize@url \@href}%
\providecommand \@href[1]{\@@startlink{#1}\@@href}%
\providecommand \@@href[1]{\endgroup#1\@@endlink}%
\providecommand \@sanitize@url [0]{\catcode `\\12\catcode `\$12\catcode
  `\&12\catcode `\#12\catcode `\^12\catcode `\_12\catcode `\%12\relax}%
\providecommand \@@startlink[1]{}%
\providecommand \@@endlink[0]{}%
\providecommand \url  [0]{\begingroup\@sanitize@url \@url }%
\providecommand \@url [1]{\endgroup\@href {#1}{\urlprefix }}%
\providecommand \urlprefix  [0]{URL }%
\providecommand \Eprint [0]{\href }%
\providecommand \doibase [0]{http://dx.doi.org/}%
\providecommand \selectlanguage [0]{\@gobble}%
\providecommand \bibinfo  [0]{\@secondoftwo}%
\providecommand \bibfield  [0]{\@secondoftwo}%
\providecommand \translation [1]{[#1]}%
\providecommand \BibitemOpen [0]{}%
\providecommand \bibitemStop [0]{}%
\providecommand \bibitemNoStop [0]{.\EOS\space}%
\providecommand \EOS [0]{\spacefactor3000\relax}%
\providecommand \BibitemShut  [1]{\csname bibitem#1\endcsname}%
\let\auto@bib@innerbib\@empty
\bibitem [{\citenamefont {Vollhardt}\ and\ \citenamefont
  {W{\"o}lfle}(1990)}]{vollhardt:book}%
  \BibitemOpen
  \bibfield  {author} {\bibinfo {author} {\bibfnamefont {D.}~\bibnamefont
  {Vollhardt}}\ and\ \bibinfo {author} {\bibfnamefont {P.}~\bibnamefont
  {W{\"o}lfle}},\ }\href@noop {} {\emph {\bibinfo {title} {The Superfluid Phase
  of Helium 3}}}\ (\bibinfo  {publisher} {Taylor and Francis},\ \bibinfo
  {address} {London},\ \bibinfo {year} {1990})\BibitemShut {NoStop}%
\bibitem [{\citenamefont {Schnyder}\ \emph {et~al.}(2008)\citenamefont
  {Schnyder}, \citenamefont {Ryu}, \citenamefont {Furusaki},\ and\
  \citenamefont {Ludwig}}]{schnyder:2008}%
  \BibitemOpen
  \bibfield  {author} {\bibinfo {author} {\bibfnamefont {A.~P.}\ \bibnamefont
  {Schnyder}}, \bibinfo {author} {\bibfnamefont {S.}~\bibnamefont {Ryu}},
  \bibinfo {author} {\bibfnamefont {A.}~\bibnamefont {Furusaki}}, \ and\
  \bibinfo {author} {\bibfnamefont {A.~W.~W.}\ \bibnamefont {Ludwig}},\ }\href
  {\doibase 10.1103/PhysRevB.78.195125} {\bibfield  {journal} {\bibinfo
  {journal} {Phys. Rev. B}\ }\textbf {\bibinfo {volume} {78}},\ \bibinfo
  {pages} {195125} (\bibinfo {year} {2008})}\BibitemShut {NoStop}%
\bibitem [{\citenamefont {Chung}\ and\ \citenamefont
  {Zhang}(2009)}]{chung:2009}%
  \BibitemOpen
  \bibfield  {author} {\bibinfo {author} {\bibfnamefont {S.~B.}\ \bibnamefont
  {Chung}}\ and\ \bibinfo {author} {\bibfnamefont {S.-C.}\ \bibnamefont
  {Zhang}},\ }\href@noop {} {\bibfield  {journal} {\bibinfo  {journal} {Phys.
  Rev. Lett.}\ }\textbf {\bibinfo {volume} {103}},\ \bibinfo {pages} {235301}
  (\bibinfo {year} {2009})}\BibitemShut {NoStop}%
\bibitem [{\citenamefont {Volovik}(2009)}]{volovik:2009a}%
  \BibitemOpen
  \bibfield  {author} {\bibinfo {author} {\bibfnamefont {G.~E.}\ \bibnamefont
  {Volovik}},\ }\href@noop {} {\bibfield  {journal} {\bibinfo  {journal}
  {Pis'ma Zh. Eksp. Teor. Fiz.}\ }\textbf {\bibinfo {volume} {90}},\ \bibinfo
  {pages} {440} (\bibinfo {year} {2009})}\BibitemShut {NoStop}%
\bibitem [{\citenamefont {Nagato}\ \emph {et~al.}(2009)\citenamefont {Nagato},
  \citenamefont {Higashitani},\ and\ \citenamefont {Nagai}}]{nagato:2009}%
  \BibitemOpen
  \bibfield  {author} {\bibinfo {author} {\bibfnamefont {Y.}~\bibnamefont
  {Nagato}}, \bibinfo {author} {\bibfnamefont {S.}~\bibnamefont {Higashitani}},
  \ and\ \bibinfo {author} {\bibfnamefont {K.}~\bibnamefont {Nagai}},\
  }\href@noop {} {\bibfield  {journal} {\bibinfo  {journal} {J. Phys. Soc.
  Jpn.}\ }\textbf {\bibinfo {volume} {78}},\ \bibinfo {pages} {123603}
  (\bibinfo {year} {2009})}\BibitemShut {NoStop}%
\bibitem [{\citenamefont {Tsutsumi}\ \emph {et~al.}(2011)\citenamefont
  {Tsutsumi}, \citenamefont {Ichioka},\ and\ \citenamefont
  {Machida}}]{tsutsumi:2011b}%
  \BibitemOpen
  \bibfield  {author} {\bibinfo {author} {\bibfnamefont {Y.}~\bibnamefont
  {Tsutsumi}}, \bibinfo {author} {\bibfnamefont {M.}~\bibnamefont {Ichioka}}, \
  and\ \bibinfo {author} {\bibfnamefont {K.}~\bibnamefont {Machida}},\ }\href
  {\doibase 10.1103/PhysRevB.83.094510} {\bibfield  {journal} {\bibinfo
  {journal} {Phys. Rev. B}\ }\textbf {\bibinfo {volume} {83}},\ \bibinfo
  {pages} {094510} (\bibinfo {year} {2011})}\BibitemShut {NoStop}%
\bibitem [{\citenamefont {Tsutsumi}\ and\ \citenamefont
  {Machida}(2012)}]{tsutsumi:2012c}%
  \BibitemOpen
  \bibfield  {author} {\bibinfo {author} {\bibfnamefont {Y.}~\bibnamefont
  {Tsutsumi}}\ and\ \bibinfo {author} {\bibfnamefont {K.}~\bibnamefont
  {Machida}},\ }\href {\doibase 10.1143/JPSJ.81.074607} {\bibfield  {journal}
  {\bibinfo  {journal} {J. Phys. Soc. Jpn.}\ }\textbf {\bibinfo {volume}
  {81}},\ \bibinfo {pages} {074607} (\bibinfo {year} {2012})}\BibitemShut
  {NoStop}%
\bibitem [{\citenamefont {Wu}\ and\ \citenamefont {Sauls}(2013)}]{wu:2013}%
  \BibitemOpen
  \bibfield  {author} {\bibinfo {author} {\bibfnamefont {H.}~\bibnamefont
  {Wu}}\ and\ \bibinfo {author} {\bibfnamefont {J.~A.}\ \bibnamefont {Sauls}},\
  }\href {\doibase 10.1103/PhysRevB.88.184506} {\bibfield  {journal} {\bibinfo
  {journal} {Phys. Rev. B}\ }\textbf {\bibinfo {volume} {88}},\ \bibinfo
  {pages} {184506} (\bibinfo {year} {2013})}\BibitemShut {NoStop}%
\bibitem [{\citenamefont {Okuda}\ and\ \citenamefont
  {Nomura}(2012)}]{okuda:2012}%
  \BibitemOpen
  \bibfield  {author} {\bibinfo {author} {\bibfnamefont {Y.}~\bibnamefont
  {Okuda}}\ and\ \bibinfo {author} {\bibfnamefont {R.}~\bibnamefont {Nomura}},\
  }\href {http://stacks.iop.org/0953-8984/24/i=34/a=343201} {\bibfield
  {journal} {\bibinfo  {journal} {J. Phys.: Condens. Matter}\ }\textbf
  {\bibinfo {volume} {24}},\ \bibinfo {pages} {343201} (\bibinfo {year}
  {2012})}\BibitemShut {NoStop}%
\bibitem [{\citenamefont {Fu}(2011)}]{fu:2011}%
  \BibitemOpen
  \bibfield  {author} {\bibinfo {author} {\bibfnamefont {L.}~\bibnamefont
  {Fu}},\ }\href {\doibase 10.1103/PhysRevLett.106.106802} {\bibfield
  {journal} {\bibinfo  {journal} {Phys. Rev. Lett.}\ }\textbf {\bibinfo
  {volume} {106}},\ \bibinfo {pages} {106802} (\bibinfo {year}
  {2011})}\BibitemShut {NoStop}%
\bibitem [{\citenamefont {Ueno}\ \emph {et~al.}(2013)\citenamefont {Ueno},
  \citenamefont {Yamakage}, \citenamefont {Tanaka},\ and\ \citenamefont
  {Sato}}]{ueno:2013}%
  \BibitemOpen
  \bibfield  {author} {\bibinfo {author} {\bibfnamefont {Y.}~\bibnamefont
  {Ueno}}, \bibinfo {author} {\bibfnamefont {A.}~\bibnamefont {Yamakage}},
  \bibinfo {author} {\bibfnamefont {Y.}~\bibnamefont {Tanaka}}, \ and\ \bibinfo
  {author} {\bibfnamefont {M.}~\bibnamefont {Sato}},\ }\href {\doibase
  10.1103/PhysRevLett.111.087002} {\bibfield  {journal} {\bibinfo  {journal}
  {Phys. Rev. Lett.}\ }\textbf {\bibinfo {volume} {111}},\ \bibinfo {pages}
  {087002} (\bibinfo {year} {2013})}\BibitemShut {NoStop}%
\bibitem [{\citenamefont {Tsutsumi}\ \emph {et~al.}(2013)\citenamefont
  {Tsutsumi}, \citenamefont {Ishikawa}, \citenamefont {Kawakami}, \citenamefont
  {Mizushima}, \citenamefont {Sato}, \citenamefont {Ichioka},\ and\
  \citenamefont {Machida}}]{tsutsumi:2013}%
  \BibitemOpen
  \bibfield  {author} {\bibinfo {author} {\bibfnamefont {Y.}~\bibnamefont
  {Tsutsumi}}, \bibinfo {author} {\bibfnamefont {M.}~\bibnamefont {Ishikawa}},
  \bibinfo {author} {\bibfnamefont {T.}~\bibnamefont {Kawakami}}, \bibinfo
  {author} {\bibfnamefont {T.}~\bibnamefont {Mizushima}}, \bibinfo {author}
  {\bibfnamefont {M.}~\bibnamefont {Sato}}, \bibinfo {author} {\bibfnamefont
  {M.}~\bibnamefont {Ichioka}}, \ and\ \bibinfo {author} {\bibfnamefont
  {K.}~\bibnamefont {Machida}},\ }\href {\doibase 10.1143/JPSJ.82.113707}
  {\bibfield  {journal} {\bibinfo  {journal} {J. Phys. Soc. Jpn.}\ }\textbf
  {\bibinfo {volume} {82}},\ \bibinfo {pages} {113707} (\bibinfo {year}
  {2013})}\BibitemShut {NoStop}%
\bibitem [{\citenamefont {Chiu}\ \emph {et~al.}(2013)\citenamefont {Chiu},
  \citenamefont {Yao},\ and\ \citenamefont {Ryu}}]{chiu:2013}%
  \BibitemOpen
  \bibfield  {author} {\bibinfo {author} {\bibfnamefont {C.-K.}\ \bibnamefont
  {Chiu}}, \bibinfo {author} {\bibfnamefont {H.}~\bibnamefont {Yao}}, \ and\
  \bibinfo {author} {\bibfnamefont {S.}~\bibnamefont {Ryu}},\ }\href {\doibase
  10.1103/PhysRevB.88.075142} {\bibfield  {journal} {\bibinfo  {journal} {Phys.
  Rev. B}\ }\textbf {\bibinfo {volume} {88}},\ \bibinfo {pages} {075142}
  (\bibinfo {year} {2013})}\BibitemShut {NoStop}%
\bibitem [{\citenamefont {Morimoto}\ and\ \citenamefont
  {Furusaki}(2013)}]{morimoto:2013}%
  \BibitemOpen
  \bibfield  {author} {\bibinfo {author} {\bibfnamefont {T.}~\bibnamefont
  {Morimoto}}\ and\ \bibinfo {author} {\bibfnamefont {A.}~\bibnamefont
  {Furusaki}},\ }\href {\doibase 10.1103/PhysRevB.88.125129} {\bibfield
  {journal} {\bibinfo  {journal} {Phys. Rev. B}\ }\textbf {\bibinfo {volume}
  {88}},\ \bibinfo {pages} {125129} (\bibinfo {year} {2013})}\BibitemShut
  {NoStop}%
\bibitem [{\citenamefont {Shiozaki}\ and\ \citenamefont
  {Sato}(2014)}]{shiozaki:2014}%
  \BibitemOpen
  \bibfield  {author} {\bibinfo {author} {\bibfnamefont {K.}~\bibnamefont
  {Shiozaki}}\ and\ \bibinfo {author} {\bibfnamefont {M.}~\bibnamefont
  {Sato}},\ }\href {\doibase 10.1103/PhysRevB.90.165114} {\bibfield  {journal}
  {\bibinfo  {journal} {Phys. Rev. B}\ }\textbf {\bibinfo {volume} {90}},\
  \bibinfo {pages} {165114} (\bibinfo {year} {2014})}\BibitemShut {NoStop}%
\bibitem [{\citenamefont {Mizushima}\ \emph {et~al.}()\citenamefont
  {Mizushima}, \citenamefont {Tsutsumi}, \citenamefont {Sato},\ and\
  \citenamefont {Machida}}]{mizushima:arXiv}%
  \BibitemOpen
  \bibfield  {author} {\bibinfo {author} {\bibfnamefont {T.}~\bibnamefont
  {Mizushima}}, \bibinfo {author} {\bibfnamefont {Y.}~\bibnamefont {Tsutsumi}},
  \bibinfo {author} {\bibfnamefont {M.}~\bibnamefont {Sato}}, \ and\ \bibinfo
  {author} {\bibfnamefont {K.}~\bibnamefont {Machida}},\ }\href@noop {}
  {\bibinfo  {journal} {arXiv:1409.6094}\ }\BibitemShut {NoStop}%
\bibitem [{\citenamefont {Mizushima}\ \emph {et~al.}(2012)\citenamefont
  {Mizushima}, \citenamefont {Sato},\ and\ \citenamefont
  {Machida}}]{mizushima:2012b}%
  \BibitemOpen
\bibfield  {journal} {  }\bibfield  {author} {\bibinfo {author} {\bibfnamefont
  {T.}~\bibnamefont {Mizushima}}, \bibinfo {author} {\bibfnamefont
  {M.}~\bibnamefont {Sato}}, \ and\ \bibinfo {author} {\bibfnamefont
  {K.}~\bibnamefont {Machida}},\ }\href {\doibase
  10.1103/PhysRevLett.109.165301} {\bibfield  {journal} {\bibinfo  {journal}
  {Phys. Rev. Lett.}\ }\textbf {\bibinfo {volume} {109}},\ \bibinfo {pages}
  {165301} (\bibinfo {year} {2012})}\BibitemShut {NoStop}%
\bibitem [{\citenamefont {Mizushima}(2012)}]{mizushima:2012}%
  \BibitemOpen
  \bibfield  {author} {\bibinfo {author} {\bibfnamefont {T.}~\bibnamefont
  {Mizushima}},\ }\href {\doibase 10.1103/PhysRevB.86.094518} {\bibfield
  {journal} {\bibinfo  {journal} {Phys. Rev. B}\ }\textbf {\bibinfo {volume}
  {86}},\ \bibinfo {pages} {094518} (\bibinfo {year} {2012})}\BibitemShut
  {NoStop}%
\bibitem [{\citenamefont {Salomaa}\ and\ \citenamefont
  {Volovik}(1987)}]{salomaa:1987}%
  \BibitemOpen
  \bibfield  {author} {\bibinfo {author} {\bibfnamefont {M.~M.}\ \bibnamefont
  {Salomaa}}\ and\ \bibinfo {author} {\bibfnamefont {G.~E.}\ \bibnamefont
  {Volovik}},\ }\href {\doibase 10.1103/RevModPhys.59.533} {\bibfield
  {journal} {\bibinfo  {journal} {Rev. Mod. Phys.}\ }\textbf {\bibinfo {volume}
  {59}},\ \bibinfo {pages} {533} (\bibinfo {year} {1987})}\BibitemShut
  {NoStop}%
\bibitem [{\citenamefont {Volovik}(2003)}]{volovik:book}%
  \BibitemOpen
  \bibfield  {author} {\bibinfo {author} {\bibfnamefont {G.~E.}\ \bibnamefont
  {Volovik}},\ }\href@noop {} {\emph {\bibinfo {title} {The Universe in a
  Helium Droplet}}}\ (\bibinfo  {publisher} {Clarendon},\ \bibinfo {address}
  {Oxford},\ \bibinfo {year} {2003})\BibitemShut {NoStop}%
\bibitem [{\citenamefont {Ohmi}\ \emph {et~al.}(1983)\citenamefont {Ohmi},
  \citenamefont {Tsuneto},\ and\ \citenamefont {Fujita}}]{ohmi:1983}%
  \BibitemOpen
  \bibfield  {author} {\bibinfo {author} {\bibfnamefont {T.}~\bibnamefont
  {Ohmi}}, \bibinfo {author} {\bibfnamefont {T.}~\bibnamefont {Tsuneto}}, \
  and\ \bibinfo {author} {\bibfnamefont {T.}~\bibnamefont {Fujita}},\ }\href
  {\doibase 10.1143/PTP.70.647} {\bibfield  {journal} {\bibinfo  {journal}
  {Prog. Theor. Phys.}\ }\textbf {\bibinfo {volume} {70}},\ \bibinfo {pages}
  {647} (\bibinfo {year} {1983})}\BibitemShut {NoStop}%
\bibitem [{\citenamefont {Hakonen}\ \emph
  {et~al.}(1983{\natexlab{a}})\citenamefont {Hakonen}, \citenamefont {Ikkala},
  \citenamefont {Islander}, \citenamefont {Lounasmaa},\ and\ \citenamefont
  {Volovik}}]{hakonen:1983}%
  \BibitemOpen
  \bibfield  {author} {\bibinfo {author} {\bibfnamefont {P.}~\bibnamefont
  {Hakonen}}, \bibinfo {author} {\bibfnamefont {O.}~\bibnamefont {Ikkala}},
  \bibinfo {author} {\bibfnamefont {S.}~\bibnamefont {Islander}}, \bibinfo
  {author} {\bibfnamefont {O.}~\bibnamefont {Lounasmaa}}, \ and\ \bibinfo
  {author} {\bibfnamefont {G.}~\bibnamefont {Volovik}},\ }\href {\doibase
  10.1007/BF00682488} {\bibfield  {journal} {\bibinfo  {journal} {J. Low Temp.
  Phys.}\ }\textbf {\bibinfo {volume} {53}},\ \bibinfo {pages} {425} (\bibinfo
  {year} {1983}{\natexlab{a}})}\BibitemShut {NoStop}%
\bibitem [{\citenamefont {Hakonen}\ \emph
  {et~al.}(1983{\natexlab{b}})\citenamefont {Hakonen}, \citenamefont {Krusius},
  \citenamefont {Salomaa}, \citenamefont {Simola}, \citenamefont {Bunkov},
  \citenamefont {Mineev},\ and\ \citenamefont {Volovik}}]{hakonen:1983b}%
  \BibitemOpen
  \bibfield  {author} {\bibinfo {author} {\bibfnamefont {P.~J.}\ \bibnamefont
  {Hakonen}}, \bibinfo {author} {\bibfnamefont {M.}~\bibnamefont {Krusius}},
  \bibinfo {author} {\bibfnamefont {M.~M.}\ \bibnamefont {Salomaa}}, \bibinfo
  {author} {\bibfnamefont {J.~T.}\ \bibnamefont {Simola}}, \bibinfo {author}
  {\bibfnamefont {Y.~M.}\ \bibnamefont {Bunkov}}, \bibinfo {author}
  {\bibfnamefont {V.~P.}\ \bibnamefont {Mineev}}, \ and\ \bibinfo {author}
  {\bibfnamefont {G.~E.}\ \bibnamefont {Volovik}},\ }\href {\doibase
  10.1103/PhysRevLett.51.1362} {\bibfield  {journal} {\bibinfo  {journal}
  {Phys. Rev. Lett.}\ }\textbf {\bibinfo {volume} {51}},\ \bibinfo {pages}
  {1362} (\bibinfo {year} {1983}{\natexlab{b}})}\BibitemShut {NoStop}%
\bibitem [{\citenamefont {Pekola}\ \emph
  {et~al.}(1984{\natexlab{a}})\citenamefont {Pekola}, \citenamefont {Simola},
  \citenamefont {Hakonen}, \citenamefont {Krusius}, \citenamefont {Lounasmaa},
  \citenamefont {Nummila}, \citenamefont {Mamniashvili}, \citenamefont
  {Packard},\ and\ \citenamefont {Volovik}}]{pekola:1984}%
  \BibitemOpen
  \bibfield  {author} {\bibinfo {author} {\bibfnamefont {J.~P.}\ \bibnamefont
  {Pekola}}, \bibinfo {author} {\bibfnamefont {J.~T.}\ \bibnamefont {Simola}},
  \bibinfo {author} {\bibfnamefont {P.~J.}\ \bibnamefont {Hakonen}}, \bibinfo
  {author} {\bibfnamefont {M.}~\bibnamefont {Krusius}}, \bibinfo {author}
  {\bibfnamefont {O.~V.}\ \bibnamefont {Lounasmaa}}, \bibinfo {author}
  {\bibfnamefont {K.~K.}\ \bibnamefont {Nummila}}, \bibinfo {author}
  {\bibfnamefont {G.}~\bibnamefont {Mamniashvili}}, \bibinfo {author}
  {\bibfnamefont {R.~E.}\ \bibnamefont {Packard}}, \ and\ \bibinfo {author}
  {\bibfnamefont {G.~E.}\ \bibnamefont {Volovik}},\ }\href {\doibase
  10.1103/PhysRevLett.53.584} {\bibfield  {journal} {\bibinfo  {journal} {Phys.
  Rev. Lett.}\ }\textbf {\bibinfo {volume} {53}},\ \bibinfo {pages} {584}
  (\bibinfo {year} {1984}{\natexlab{a}})}\BibitemShut {NoStop}%
\bibitem [{\citenamefont {Pekola}\ \emph
  {et~al.}(1984{\natexlab{b}})\citenamefont {Pekola}, \citenamefont {Simola},
  \citenamefont {Nummila}, \citenamefont {Lounasmaa},\ and\ \citenamefont
  {Packard}}]{pekola:1984b}%
  \BibitemOpen
  \bibfield  {author} {\bibinfo {author} {\bibfnamefont {J.~P.}\ \bibnamefont
  {Pekola}}, \bibinfo {author} {\bibfnamefont {J.~T.}\ \bibnamefont {Simola}},
  \bibinfo {author} {\bibfnamefont {K.~K.}\ \bibnamefont {Nummila}}, \bibinfo
  {author} {\bibfnamefont {O.~V.}\ \bibnamefont {Lounasmaa}}, \ and\ \bibinfo
  {author} {\bibfnamefont {R.~E.}\ \bibnamefont {Packard}},\ }\href {\doibase
  10.1103/PhysRevLett.53.70} {\bibfield  {journal} {\bibinfo  {journal} {Phys.
  Rev. Lett.}\ }\textbf {\bibinfo {volume} {53}},\ \bibinfo {pages} {70}
  (\bibinfo {year} {1984}{\natexlab{b}})}\BibitemShut {NoStop}%
\bibitem [{\citenamefont {Pekola}\ and\ \citenamefont
  {Simola}(1985)}]{pekola:1985}%
  \BibitemOpen
  \bibfield  {author} {\bibinfo {author} {\bibfnamefont {J.}~\bibnamefont
  {Pekola}}\ and\ \bibinfo {author} {\bibfnamefont {J.}~\bibnamefont
  {Simola}},\ }\href@noop {} {\bibfield  {journal} {\bibinfo  {journal} {J. Low
  Temp. Phys.}\ }\textbf {\bibinfo {volume} {58}},\ \bibinfo {pages} {555}
  (\bibinfo {year} {1985})}\BibitemShut {NoStop}%
\bibitem [{\citenamefont {Thuneberg}(1986)}]{thuneberg:1986b}%
  \BibitemOpen
  \bibfield  {author} {\bibinfo {author} {\bibfnamefont {E.~V.}\ \bibnamefont
  {Thuneberg}},\ }\href {\doibase 10.1103/PhysRevLett.56.359} {\bibfield
  {journal} {\bibinfo  {journal} {Phys. Rev. Lett.}\ }\textbf {\bibinfo
  {volume} {56}},\ \bibinfo {pages} {359} (\bibinfo {year} {1986})}\BibitemShut
  {NoStop}%
\bibitem [{\citenamefont {Thuneberg}(1987)}]{thuneberg:1987}%
  \BibitemOpen
  \bibfield  {author} {\bibinfo {author} {\bibfnamefont {E.~V.}\ \bibnamefont
  {Thuneberg}},\ }\href@noop {} {\bibfield  {journal} {\bibinfo  {journal}
  {Phys. Rev. B}\ }\textbf {\bibinfo {volume} {36}},\ \bibinfo {pages} {3583}
  (\bibinfo {year} {1987})}\BibitemShut {NoStop}%
\bibitem [{\citenamefont {Kondo}\ \emph {et~al.}(1991)\citenamefont {Kondo},
  \citenamefont {Korhonen}, \citenamefont {Krusius}, \citenamefont {Dmitriev},
  \citenamefont {Mukharsky}, \citenamefont {Sonin},\ and\ \citenamefont
  {Volovik}}]{kondo:1991}%
  \BibitemOpen
  \bibfield  {author} {\bibinfo {author} {\bibfnamefont {Y.}~\bibnamefont
  {Kondo}}, \bibinfo {author} {\bibfnamefont {J.~S.}\ \bibnamefont {Korhonen}},
  \bibinfo {author} {\bibfnamefont {M.}~\bibnamefont {Krusius}}, \bibinfo
  {author} {\bibfnamefont {V.~V.}\ \bibnamefont {Dmitriev}}, \bibinfo {author}
  {\bibfnamefont {Y.~M.}\ \bibnamefont {Mukharsky}}, \bibinfo {author}
  {\bibfnamefont {E.~B.}\ \bibnamefont {Sonin}}, \ and\ \bibinfo {author}
  {\bibfnamefont {G.~E.}\ \bibnamefont {Volovik}},\ }\href@noop {} {\bibfield
  {journal} {\bibinfo  {journal} {Phys. Rev. Lett.}\ }\textbf {\bibinfo
  {volume} {67}},\ \bibinfo {pages} {81} (\bibinfo {year} {1991})}\BibitemShut
  {NoStop}%
\bibitem [{\citenamefont {Silaev}(2009)}]{silaev:2009}%
  \BibitemOpen
  \bibfield  {author} {\bibinfo {author} {\bibfnamefont {M.~A.}\ \bibnamefont
  {Silaev}},\ }\href@noop {} {\bibfield  {journal} {\bibinfo  {journal} {JETP
  Lett.}\ }\textbf {\bibinfo {volume} {90}},\ \bibinfo {pages} {391} (\bibinfo
  {year} {2009})}\BibitemShut {NoStop}%
\bibitem [{\citenamefont {Caroli}\ \emph {et~al.}(1964)\citenamefont {Caroli},
  \citenamefont {de~Gennes},\ and\ \citenamefont {Matricon}}]{caroli:1964}%
  \BibitemOpen
  \bibfield  {author} {\bibinfo {author} {\bibfnamefont {C.}~\bibnamefont
  {Caroli}}, \bibinfo {author} {\bibfnamefont {P.}~\bibnamefont {de~Gennes}}, \
  and\ \bibinfo {author} {\bibfnamefont {J.}~\bibnamefont {Matricon}},\
  }\href@noop {} {\bibfield  {journal} {\bibinfo  {journal} {Phys. Lett.}\
  }\textbf {\bibinfo {volume} {9}},\ \bibinfo {pages} {307} (\bibinfo {year}
  {1964})}\BibitemShut {NoStop}%
\bibitem [{\citenamefont {Kawakami}\ \emph {et~al.}(2011)\citenamefont
  {Kawakami}, \citenamefont {Mizushima},\ and\ \citenamefont
  {Machida}}]{kawakami:2011}%
  \BibitemOpen
  \bibfield  {author} {\bibinfo {author} {\bibfnamefont {T.}~\bibnamefont
  {Kawakami}}, \bibinfo {author} {\bibfnamefont {T.}~\bibnamefont {Mizushima}},
  \ and\ \bibinfo {author} {\bibfnamefont {K.}~\bibnamefont {Machida}},\ }\href
  {\doibase 10.1143/JPSJ.80.044603} {\bibfield  {journal} {\bibinfo  {journal}
  {J. Phys. Soc. Jpn.}\ }\textbf {\bibinfo {volume} {80}},\ \bibinfo {pages}
  {044603} (\bibinfo {year} {2011})}\BibitemShut {NoStop}%
\bibitem [{\citenamefont {Kaneko}\ \emph {et~al.}(2012)\citenamefont {Kaneko},
  \citenamefont {Matsuba}, \citenamefont {Hafiz}, \citenamefont {Yamasaki},
  \citenamefont {Kakizaki}, \citenamefont {Nishida}, \citenamefont {Takeya},
  \citenamefont {Hirata}, \citenamefont {Kawakami}, \citenamefont {Mizushima},\
  and\ \citenamefont {Machida}}]{kaneko:2012}%
  \BibitemOpen
  \bibfield  {author} {\bibinfo {author} {\bibfnamefont {S.}~\bibnamefont
  {Kaneko}}, \bibinfo {author} {\bibfnamefont {K.}~\bibnamefont {Matsuba}},
  \bibinfo {author} {\bibfnamefont {M.}~\bibnamefont {Hafiz}}, \bibinfo
  {author} {\bibfnamefont {K.}~\bibnamefont {Yamasaki}}, \bibinfo {author}
  {\bibfnamefont {E.}~\bibnamefont {Kakizaki}}, \bibinfo {author}
  {\bibfnamefont {N.}~\bibnamefont {Nishida}}, \bibinfo {author} {\bibfnamefont
  {H.}~\bibnamefont {Takeya}}, \bibinfo {author} {\bibfnamefont
  {K.}~\bibnamefont {Hirata}}, \bibinfo {author} {\bibfnamefont
  {T.}~\bibnamefont {Kawakami}}, \bibinfo {author} {\bibfnamefont
  {T.}~\bibnamefont {Mizushima}}, \ and\ \bibinfo {author} {\bibfnamefont
  {K.}~\bibnamefont {Machida}},\ }\href {\doibase 10.1143/JPSJ.81.063701}
  {\bibfield  {journal} {\bibinfo  {journal} {J. Phys. Soc. Jpn.}\ }\textbf
  {\bibinfo {volume} {81}},\ \bibinfo {pages} {063701} (\bibinfo {year}
  {2012})}\BibitemShut {NoStop}%
\bibitem [{\citenamefont {Mizushima}\ and\ \citenamefont
  {Machida}(2010{\natexlab{a}})}]{mizushima:2010b}%
  \BibitemOpen
  \bibfield  {author} {\bibinfo {author} {\bibfnamefont {T.}~\bibnamefont
  {Mizushima}}\ and\ \bibinfo {author} {\bibfnamefont {K.}~\bibnamefont
  {Machida}},\ }\href {\doibase 10.1103/PhysRevA.82.023624} {\bibfield
  {journal} {\bibinfo  {journal} {Phys. Rev. A}\ }\textbf {\bibinfo {volume}
  {82}},\ \bibinfo {pages} {023624} (\bibinfo {year}
  {2010}{\natexlab{a}})}\BibitemShut {NoStop}%
\bibitem [{\citenamefont {Sato}\ \emph {et~al.}(2014)\citenamefont {Sato},
  \citenamefont {Yamakage},\ and\ \citenamefont {Mizushima}}]{sato:2014}%
  \BibitemOpen
  \bibfield  {author} {\bibinfo {author} {\bibfnamefont {M.}~\bibnamefont
  {Sato}}, \bibinfo {author} {\bibfnamefont {A.}~\bibnamefont {Yamakage}}, \
  and\ \bibinfo {author} {\bibfnamefont {T.}~\bibnamefont {Mizushima}},\ }\href
  {\doibase http://dx.doi.org/10.1016/j.physe.2013.07.011} {\bibfield
  {journal} {\bibinfo  {journal} {Physica E}\ }\textbf {\bibinfo {volume}
  {55}},\ \bibinfo {pages} {20} (\bibinfo {year} {2014})}\BibitemShut {NoStop}%
\bibitem [{\citenamefont {Eilenberger}(1968)}]{eilenberger:1968}%
  \BibitemOpen
  \bibfield  {author} {\bibinfo {author} {\bibfnamefont {G.}~\bibnamefont
  {Eilenberger}},\ }\href@noop {} {\bibfield  {journal} {\bibinfo  {journal}
  {Z. Phys.}\ }\textbf {\bibinfo {volume} {214}},\ \bibinfo {pages} {195}
  (\bibinfo {year} {1968})}\BibitemShut {NoStop}%
\bibitem [{\citenamefont {Serene}\ and\ \citenamefont
  {Rainer}(1983)}]{serene:1983}%
  \BibitemOpen
  \bibfield  {author} {\bibinfo {author} {\bibfnamefont {J.~W.}\ \bibnamefont
  {Serene}}\ and\ \bibinfo {author} {\bibfnamefont {D.}~\bibnamefont
  {Rainer}},\ }\href@noop {} {\bibfield  {journal} {\bibinfo  {journal} {Phys.
  Rep.}\ }\textbf {\bibinfo {volume} {101}},\ \bibinfo {pages} {221} (\bibinfo
  {year} {1983})}\BibitemShut {NoStop}%
\bibitem [{\citenamefont {Fogelstr{\"o}m}\ and\ \citenamefont
  {Kurkij{\"a}rvi}(1995)}]{fogelstrom:1995}%
  \BibitemOpen
  \bibfield  {author} {\bibinfo {author} {\bibfnamefont {M.}~\bibnamefont
  {Fogelstr{\"o}m}}\ and\ \bibinfo {author} {\bibfnamefont {J.}~\bibnamefont
  {Kurkij{\"a}rvi}},\ }\href@noop {} {\bibfield  {journal} {\bibinfo  {journal}
  {J. Low Temp. Phys.}\ }\textbf {\bibinfo {volume} {98}},\ \bibinfo {pages}
  {195} (\bibinfo {year} {1995})}\BibitemShut {NoStop}%
\bibitem [{\citenamefont {Gurarie}\ and\ \citenamefont
  {Radzihovsky}(2007)}]{gurarie:2007}%
  \BibitemOpen
  \bibfield  {author} {\bibinfo {author} {\bibfnamefont {V.}~\bibnamefont
  {Gurarie}}\ and\ \bibinfo {author} {\bibfnamefont {L.}~\bibnamefont
  {Radzihovsky}},\ }\href {\doibase 10.1103/PhysRevB.75.212509} {\bibfield
  {journal} {\bibinfo  {journal} {Phys. Rev. B}\ }\textbf {\bibinfo {volume}
  {75}},\ \bibinfo {pages} {212509} (\bibinfo {year} {2007})}\BibitemShut
  {NoStop}%
\bibitem [{\citenamefont {Mizushima}\ and\ \citenamefont
  {Machida}(2010{\natexlab{b}})}]{mizushima:2010}%
  \BibitemOpen
  \bibfield  {author} {\bibinfo {author} {\bibfnamefont {T.}~\bibnamefont
  {Mizushima}}\ and\ \bibinfo {author} {\bibfnamefont {K.}~\bibnamefont
  {Machida}},\ }\href {\doibase 10.1103/PhysRevA.81.053605} {\bibfield
  {journal} {\bibinfo  {journal} {Phys. Rev. A}\ }\textbf {\bibinfo {volume}
  {81}},\ \bibinfo {pages} {053605} (\bibinfo {year}
  {2010}{\natexlab{b}})}\BibitemShut {NoStop}%
\bibitem [{\citenamefont {Sato}\ and\ \citenamefont
  {Fujimoto}(2009)}]{sato:2009c}%
  \BibitemOpen
  \bibfield  {author} {\bibinfo {author} {\bibfnamefont {M.}~\bibnamefont
  {Sato}}\ and\ \bibinfo {author} {\bibfnamefont {S.}~\bibnamefont
  {Fujimoto}},\ }\href {\doibase 10.1103/PhysRevB.79.094504} {\bibfield
  {journal} {\bibinfo  {journal} {Phys. Rev. B}\ }\textbf {\bibinfo {volume}
  {79}},\ \bibinfo {pages} {094504} (\bibinfo {year} {2009})}\BibitemShut
  {NoStop}%
\bibitem [{\citenamefont {Sato}\ \emph {et~al.}(2011)\citenamefont {Sato},
  \citenamefont {Tanaka}, \citenamefont {Yada},\ and\ \citenamefont
  {Yokoyama}}]{sato:2011}%
  \BibitemOpen
  \bibfield  {author} {\bibinfo {author} {\bibfnamefont {M.}~\bibnamefont
  {Sato}}, \bibinfo {author} {\bibfnamefont {Y.}~\bibnamefont {Tanaka}},
  \bibinfo {author} {\bibfnamefont {K.}~\bibnamefont {Yada}}, \ and\ \bibinfo
  {author} {\bibfnamefont {T.}~\bibnamefont {Yokoyama}},\ }\href {\doibase
  10.1103/PhysRevB.83.224511} {\bibfield  {journal} {\bibinfo  {journal} {Phys.
  Rev. B}\ }\textbf {\bibinfo {volume} {83}},\ \bibinfo {pages} {224511}
  (\bibinfo {year} {2011})}\BibitemShut {NoStop}%
\bibitem [{\citenamefont {Qi}\ \emph {et~al.}(2008)\citenamefont {Qi},
  \citenamefont {Hughes},\ and\ \citenamefont {Zhang}}]{qi:2008}%
  \BibitemOpen
  \bibfield  {author} {\bibinfo {author} {\bibfnamefont {X.-L.}\ \bibnamefont
  {Qi}}, \bibinfo {author} {\bibfnamefont {T.~L.}\ \bibnamefont {Hughes}}, \
  and\ \bibinfo {author} {\bibfnamefont {S.-C.}\ \bibnamefont {Zhang}},\ }\href
  {\doibase 10.1103/PhysRevB.78.195424} {\bibfield  {journal} {\bibinfo
  {journal} {Phys. Rev. B}\ }\textbf {\bibinfo {volume} {78}},\ \bibinfo
  {pages} {195424} (\bibinfo {year} {2008})}\BibitemShut {NoStop}%
\bibitem [{\citenamefont {Teo}\ and\ \citenamefont {Kane}(2010)}]{teo:2010}%
  \BibitemOpen
  \bibfield  {author} {\bibinfo {author} {\bibfnamefont {J.~C.~Y.}\
  \bibnamefont {Teo}}\ and\ \bibinfo {author} {\bibfnamefont {C.~L.}\
  \bibnamefont {Kane}},\ }\href {\doibase 10.1103/PhysRevB.82.115120}
  {\bibfield  {journal} {\bibinfo  {journal} {Phys. Rev. B}\ }\textbf {\bibinfo
  {volume} {82}},\ \bibinfo {pages} {115120} (\bibinfo {year}
  {2010})}\BibitemShut {NoStop}%
\bibitem [{\citenamefont {Kitaev}(2009)}]{kitaev:2009}%
  \BibitemOpen
  \bibfield  {author} {\bibinfo {author} {\bibfnamefont {A.}~\bibnamefont
  {Kitaev}},\ }\href {\doibase http://dx.doi.org/10.1063/1.3149495} {\bibfield
  {journal} {\bibinfo  {journal} {AIP Conf. Proc.}\ }\textbf {\bibinfo {volume}
  {1134}},\ \bibinfo {pages} {22} (\bibinfo {year} {2009})}\BibitemShut
  {NoStop}%
\bibitem [{\citenamefont {Furusaki}(1999)}]{furusaki:1999}%
  \BibitemOpen
  \bibfield  {author} {\bibinfo {author} {\bibfnamefont {A.}~\bibnamefont
  {Furusaki}},\ }\href@noop {} {\bibfield  {journal} {\bibinfo  {journal}
  {Superlattices and Microstructures}\ }\textbf {\bibinfo {volume} {25}},\
  \bibinfo {pages} {809} (\bibinfo {year} {1999})}\BibitemShut {NoStop}%
\bibitem [{\citenamefont {Kashiwaya}\ and\ \citenamefont
  {Tanaka}(2000)}]{kashiwaya:2000}%
  \BibitemOpen
  \bibfield  {author} {\bibinfo {author} {\bibfnamefont {S.}~\bibnamefont
  {Kashiwaya}}\ and\ \bibinfo {author} {\bibfnamefont {Y.}~\bibnamefont
  {Tanaka}},\ }\href {http://stacks.iop.org/0034-4885/63/i=10/a=202} {\bibfield
   {journal} {\bibinfo  {journal} {Rep. Prog. Phys.}\ }\textbf {\bibinfo
  {volume} {63}},\ \bibinfo {pages} {1641} (\bibinfo {year}
  {2000})}\BibitemShut {NoStop}%
\end{thebibliography}%

\end{document}